\documentclass[11pt]{article}

\pdfoutput=1
\usepackage{fullpage}
\usepackage{latexsym}
\usepackage[scaled]{helvet}
\usepackage{times}
\usepackage{xspace}
\usepackage{graphicx}
\usepackage{subfig}
\usepackage{float}
\usepackage{amssymb,amsmath,amsfonts,amsthm}
\usepackage{mathrsfs}
\usepackage{color}
\usepackage{mathtools}
\usepackage{tabularx}
\usepackage{framed}
\usepackage{enumerate}
\usepackage{subfig}

\newcommand{\RR}{\mathbb{R}}
\newtheorem{theorem}{Theorem}
\newtheorem{lemma}[theorem]{Lemma}

\newtheorem{definition}[theorem]{Definition}

\newtheorem{observation}{Observation}
\newtheorem{claim}[theorem]{Claim}

\newcommand{\eat}[1]{}

\newcommand{\poa}{{\sc PoA}\xspace}
\newcommand{\pos}{{\sc PoS}\xspace}
\newcommand{\nash}{{\sc NE}\xspace}
\newcommand{\ep}{{\sc eq-p}\xspace}
\newcommand{\nep}{{\sc non-eq-p}\xspace}
\newcommand{\mst}{{\sc mst}\xspace}

\newcommand{\equilibrium}{{\sc balanced-equilibrium}\xspace}
\newcommand{\amortized}{{\sc balanced}\xspace}
\newcommand{\leaf}{{\sc leaf-unbalanced}\xspace}
\newcommand{\nonleaf}{{\sc non-leaf-unbalanced}\xspace}
\newcommand{\stm}{{\sc select tree move}\xspace}

\newcommand{\opt}{{\sc opt}\xspace}

\newcommand{\OPT}{\mathop{\operatorname{OPT}}}

\newcommand{\N}{N}

\title{Timing Matters: Online Dynamics in Broadcast Games\thanks{Part of this work was done when all the authors were visiting Microsoft Research - Redmond.}}
\author{
Shuchi Chawla\thanks{Computer Sciences Department, University of
  Wisconsin - Madison. Email: {\tt shuchi@cs.wisc.edu}. This work was
  supported in part by NSF grants CCF-1101429 and CCF-1320854.}
\and Joseph (Seffi) Naor\thanks{Department of Computer Science, Technion. Email: {\tt naor@cs.technion.ac.il}.}
\and Debmalya Panigrahi\thanks{Department of Computer Science, Duke University. Email: {\tt debmalya@cs.duke.edu}. This work was supported in part by NSF Awards CCF-1527084 and CCF-1535972, a Google Faculty Research Award, and a Yahoo FREP Award.}
\and Mohit Singh\thanks{Microsoft Research - Redmond. Email: {\tt mohits@microsoft.com}.}
\and Seeun William Umboh\thanks{Department of Mathematics and Computer
  Science, Eindhoven University of Technology, Netherlands. Email:
  {\tt seeun.umboh@gmail.com}. This work was supported in part by ERC
  consolidator grant 617951 and NSF grant CCF-1320854. Part of this work was done while a student at the University of Wisconsin - Madison, and while visiting the Simons Institute for the Theory of Computing.}
}
\date{}

\begin{document}

\maketitle
\begin{abstract}
  A central question in algorithmic game theory is to measure the
  inefficiency (ratio of costs) of Nash equilibria (\nash) with
  respect to socially optimal solutions. The two established metrics
  used for this purpose are price of anarchy (\poa) and price of
  stability (\pos), which respectively provide upper and lower bounds
  on this ratio. A deficiency of these metrics, however, is that they
  are purely existential and shed no light on which of the equilibrium
  states are reachable in an actual game, i.e., via natural game
  dynamics. This is particularly striking if these metrics differ
  significantly in value, such as in network design games where the
  exponential gap between the best and worst \nash states originally
  prompted the notion of \pos in game theory (Anshelevich {\em et
    al.}, FOCS 2002). In this paper, we make progress toward bridging
  this gap by studying network design games under natural game
  dynamics.

First we show that in a completely decentralized setting, where
agents arrive, depart, and make improving moves in an arbitrary order,
the inefficiency of \nash attained can be polynomially large. To the
best of our knowledge, this is the first demonstration of an \nash
with polynomial inefficiency that can actually be attained starting at
an empty state. This negative result implies that the game designer
must have some control over the interleaving of these events
(arrivals, departures, and moves) in order to force the game to attain
efficient \nash. We complement our negative result by showing that if
the game designer is allowed to execute a sequence of improving moves
to create an equilibrium state after every batch of agent arrivals or
departures, then the resulting equilibrium states attained by the game
are exponentially more efficient, i.e., the ratio of costs compared to
the optimum is only logarithmic. This result is obtained by a careful
dual charging argument where the goal of the improving moves executed
by the game designer is to dissipate the excessive charge accumulated
on any region of the network by the previous phase of agent arrivals
or departures. Overall, our two results establish that in network
games, the efficiency of equilibrium states is dictated by whether
agents are allowed to join or leave the game in arbitrary states, an
observation that might be useful in analyzing the dynamics of other
classes of games with divergent \pos and \poa bounds.
\end{abstract}
\pagenumbering{gobble}
\clearpage
\pagenumbering{arabic}
\section{Introduction}
\label{sec:introduction}

In multi-agent systems where different agents have competing objectives, it is well-known that selfish behavior can lead to suboptimal system performance. A natural question is to quantify how the system performs at a stable state or equilibrium that is consistent with selfish interests of users, relative to an optimal solution designed by a central authority. The \emph{Price of Anarchy} (\poa), that captures the relative performance of the worst possible equilibrium, and the {\em Price of Stability} (\pos), that captures the relative performance of the best possible equilibrium, are two successful and widely applied concepts developed to address this question. In a scenario where these two measures are close to each other, they provide a satisfactory resolution to understanding the quality of stable states the system is expected to reach. On the other hand, when these two measures differ significantly, the system exhibits multiple equilibria with highly varying performance, and they do not adequately address whether or not good performance would be achieved. A natural direction is to understand the quality of equilibria that can be reached organically by the agents via some dynamics. More generally, what is the minimal guidance put in place by a central authority so as to guarantee that the relative quality of the equilibrium reached is close to the best possible, that is, the price of stability?


We study these questions on \emph{broadcast games} which form a subclass of a more general class of congestion
games called {\em network design games} that were introduced by
Anshelevich~{\em et al.}~\cite{AnshelevichDKTWR08}. In a broadcast game, we are given a rooted undirected graph with costs on the edges. Every vertex has an agent residing on it and the goal of the agent is to select a path to connect to the root. The cost of the edges thus selected must be paid collectively by the agents using those edges. The Shapley cost sharing scheme stipulates that the cost of each edge is divided equally among all the agents using that edge; an agent's total {\em shared cost} is the sum of her cost share on the edges along her selected path. A state of the system, i.e., a {\em routing solution}, is defined by a collection of paths, one for each agent. The state is in Nash equilibrium (or \nash) if no agent can lower her shared cost by unilaterally changing her routing path. The existence of  \nash in broadcast
games is proved through a potential function argument, originally given by Rosenthal \cite{Rosenthal73,Monderer96}. The social cost of a solution is the sum of the costs of all edges contained in it. Observe that the socially optimal solution is a minimum spanning tree (\mst) of the graph, and thus both the \poa and \pos are defined relative to its cost.

Anshelevich~{\em et al.}~\cite{AnshelevichDKTWR08} observed that there exist instances of the broadcast game with equilibria whose cost differs from the optimum by a factor of $\Omega(n)$, where $n$ denotes the size of the graph. In other words, the \poa of the game can be as large as $\Omega(n)$. However, examples achieving this lower bound are somewhat artificial for two reasons. First, there exist alternative \nash states in every example that are much more efficient: following a long line of work \cite{FiatKLOS06,LeeL13,Li09}, Bilo {\em et al.}~\cite{BiloFM13} showed that every instance of the game contains an equilibrium whose cost is within a constant factor of the optimum, i.e., the \pos is $O(1)$\footnote{Appendix~\ref{app:examples} provides examples illustrating these bounds.}. This implies that broadcast games display multiple equilibria of widely varying quality. Second, there are no known natural (e.g., best response) dynamics that lead to these inefficient
\nash states. Given that we know of the existence of both efficient and inefficient equilibria, an intriguing question is which of these would be attained via natural game
dynamics. For example, if starting from an empty graph agents are allowed to enter the game and choose their strategies sequentially, what can we say about the quality of equilibria that emerge in such situations?


\medskip
\noindent
{\bf Our Results.}
We consider the evolution of the state in a broadcast game under the following dynamics. Starting with an empty graph, we allow the following events to modify the routing solution.
\begin{itemize}
	\item {\bf Arrival:} A set of new agents joins the game; each new agent chooses her best response (least shared cost) path given the current solution. 
	\item {\bf Departure:} A subset of existing agents leaves the game. 
	\item {\bf Move:} An agent changes her routing path to decrease her shared cost.
		We will call this an {\em improving move} (or if there is no scope of confusion,
		simply a move).
\end{itemize}
Our goal is to understand whether, and under what assumptions, the system can reach good quality (i.e., low social cost) equilibrium states. We assume that edge costs satisfy the triangle inequality and compare the cost of the equilibrium reached to \opt, defined to be the \mst of all the vertices in the graph\footnote{Observe that because we allow agents to leave the game, at the end of a sequence of moves, many vertices in the graph may not have any active agents residing at them. A natural goal then is to argue that the cost of the equilibrium reached at the end of the sequence is comparable to the cost of the minimum Steiner tree over all vertices with an active agent. However, even in extremely simple graphs, the gap between the cost of the equilibrium and the minimum Steiner tree can be as large as $\Omega(k)$, where $k$ is the number of terminals active in the end. See Appendix~\ref{app:examples} for an example. A more reasonable comparison, then, is against the minimum cost tree spanning all vertices that ever contained an active agent.}.

We first show that if the central authority does not place any restrictions on the dynamics by which the game evolves, then the equilibrium reached can be significantly worse off than the social optimum.
\begin{theorem}
\label{thm:nep}
For any large enough integer $n$, there exists an instance of the broadcast game with $n$ vertices and a sequence of arrivals and departures that terminates in an \nash of cost $\Omega(n^{1/3})$ times that of the minimum spanning tree on all the vertices.
\end{theorem}

A crucial feature of the instances we construct for the proof of Theorem~\ref{thm:nep} is that the dynamics consists only of arrivals and departures, with no improving moves in between. Although intermediate states are far from being in equilibrium (i.e., many agents want to change their paths), no improving moves are allowed until the sequence of arrivals and departures ends. When all of the arrivals and departures are done, the resulting final state is in equilibrium with significantly higher cost than the social optimum.

This lower bound shows that the central authority must have some control over the arrival/departure
events in order to ensure good quality equilibrium states. In fact,
we show a sharp dependence of the efficiency of \nash states reachable via the above dynamics
on the {\em timing} of the arrival and departure events.
In particular, we consider the following dynamics:
if an arrival or departure event moves the system out of equilibrium,
the central authority is allowed to restore equilibrium through a sequence of improving moves
before the next batch of arrivals/departures happens.
The sequence of arrivals and departures is otherwise allowed to be arbitrary, indeed adversarial,
as in the lower bound instance. We call this dynamics {\em equilibrium-preserving} (\ep) and show the following result.
\begin{theorem}
\label{thm:ep}
	For every instance of the broadcast game using \ep dynamics, the system converges to
	\nash of cost $O(\log n)$ times that of the minimum spanning tree on all the vertices.
\end{theorem}




\subsection{Our Techniques}
We first outline our techniques for the upper bound (Theorem~\ref{thm:ep}), which is our main technical result. At a very high level, our argument relies on structural properties of states reachable in \ep dynamics that we prove via induction. We then employ a charging argument for the cost of any such state against a family of dual solutions for the minimum spanning tree over the underlying graph.

Our main construct for charging the cost of a solution is a family of $O(\log n)$ dual solutions\footnote{Each dual solution serves as a lower bound on the \mst.} for the minimum spanning tree. The $j$th dual in the family is a partition of the vertices into subsets of diameter (roughly) $2^j$ whose ``centers'' are at a distance of (roughly) $2^j$ from one another; each such subset is called a level $j$ cut. For any state of the system where the routing paths form a tree, we make every vertex in the tree ``responsible'' for the first edge on its unique tree path to the root. We charge the cost of an edge of length $\ell$ to a cut of diameter roughly $\ell/4$, namely the cut in level $\lfloor\log(\ell/4)\rfloor$ that contains the vertex responsible for that edge. Our main goal is to show that for any equilibrium reachable via \ep dynamics, each cut in the dual family is charged at most once; we call such an equilibrium a ``balanced equilibrium''. A simple accounting then bounds the cost of such an equilibrium within $O(\log n)$ of the cost of the \mst.

To get a feel for how we might maintain such a balanced charging, consider a simple case where we start from a balanced equilibrium and an agent arrives at a new vertex $v$. The agent picks a path to connect to the root that, without loss of generality, consists of a new edge from $v$ to a vertex, say $w$, in the existing routing tree, and the tree path from $w$ to the root. The new edge potentially places an extra charge on a cut $C$ that was already being charged by a different vertex, say $u$, previously. Thus, our charging is out of balance. We show that it must be the case that $u$ benefits from changing its current routing path to the path obtained by using edge $(u,v)$ followed by $v$'s new path to the root: while $u$ must now pay the cost of unilaterally using edge $(u,v)$, roughly speaking, this is offset by the savings it obtains by only paying half the cost of $v$'s new edge $(v,w)$. Vertex $u$'s descendants in the routing tree likewise follow $u$ to the new path; we collectively call this sequence of improving moves a \emph{tree-follow} move. This fixes the issue of overcharging of the cut $C$. Of course, if $u$ moves to this new path, it creates a new edge $(u,v)$, which in turn potentially overcharges a different cut $C'$. We then repeat the above argument for $C'$, and so on, till a balanced state is restored.

The key invariant in the above argument is that only one cut is overcharged at a time. If multiple new arrivals happen all at once, however, this invariant no longer holds. Nevertheless, we can argue that any overcharging is done by leaf nodes, i.e. nodes that do not have any descendants in the routing tree because they are freshly arrived. This observation allows us to argue as before that for any cut that is overcharged, one of the vertices charging it has an improving move that removes this extra charge. We can now distinguish between states of the routing tree that are ``balanced'', and those that are ``leaf-unbalanced'', meaning in the latter case that every cut is charged by at most one non-leaf, but potentially many leaves. From leaf-unbalanced states we carry out a carefully ordered sequence of improving moves with the goal of eventually reaching a balanced state. Unfortunately, some of these moves apply to non-leaves, and can lead to a state where a cut is charged simultaneously by two non-leaf vertices. We show, however, that in any state reachable within \ep dynamics, at most one cut can be charged by multiple non-leaf vertices, and never by more than two non-leaf vertices. We call such states ``non-leaf-unbalanced''.

The departure of one or more agents does not affect the charging of cuts, but may lead to the introduction of Steiner vertices. We continue to hold Steiner vertices responsible for the first edge on their tree path to the root. This does not create a problem, as our family of dual solutions covers the entire graph and its total cost is bounded against the cost of the \mst rather than a minimum Steiner tree.

Putting everything together, we argue that \ep dynamics cycles through four types of states -- non-leaf-unbalanced, leaf-unbalanced, balanced, and balanced-equilibrium. In states of the first three types, we can always find an improving move leading to one of the four types of states. Each improving move leads to a decrease in the standard potential function for the game \cite{Rosenthal73} (discussed in more detail in the following section). Therefore, the sequence of moves terminates within a finite number of steps at a balanced-equilibrium state. The cost of this equilibrium can then be charged against the family of dual solutions described above.

For our lower bound for \nep dynamics (Theorem~\ref{thm:nep}), the high level idea is to create a \poa type of instance in which multiple different agents, that are located much closer to each other relative to the root, nevertheless follow independent paths to the root in the final solution. Such a solution can be made stable by ensuring that each agent is co-located with a large group of other agents. Call this entire set of agents the primary agents in the game, and the corresponding vertices the primary vertices. One challenge with creating such a state dynamically is that after we have placed an agent at one of the primary vertices in the graph, when we place an agent at another close-by primary vertex, the best path for this agent is to take the short-cut to the first agent, and follow the latter's path to the root, thereby saving on cost via sharing. In order to get around this, and force every agent to take an independent path to the root, we introduce many new ``auxiliary'' agents at intermediate vertices along the desired path so as to make this path look cheap. We then remove these auxiliary agents so as to continue the process of introducing new primary agents at close-by vertices. The second challenge that arises is to guarantee that the introduction of auxiliary agents does not increase the cost of the \mst by too much. In other words, while the collection of independent paths for the primary agents should have a large total cost, there exists an \mst covering all intermediate vertices on all of the independent paths at a much lower cost. We achieve this by interconnecting the independent paths in such a manner that these paths successively converge and diverge from each other in a zig-zag fashion from the root to the primary vertices. We present this construction in Appendix~\ref{sec:nep}. 

\subsection{Related Work}

Broadcast games form a subclass of a more general class of congestion
games called {\em network design games} that were introduced by
Anshelevich~{\em et al.}~\cite{AnshelevichDKTWR08}. The existence of \nash in such
games is guaranteed by the fact that all congestion games are also potential games \cite{Rosenthal73,Monderer96}.
Substantial research effort in the last
decade has been spent on bounding the {\em price of stability} (\pos) of broadcast games.
The potential function of Rosenthal \cite{Rosenthal73}, originally used to show the existence of \nash, was also used
to prove that the \pos is at most $O(\log n)$~\cite{AnshelevichDKTWR08}, which is already
an exponential improvement over the \poa bound. The \pos bound was subsequently improved
to $O(\log \log n)$ by Fiat~{\em et al.}~\cite{FiatKLOS06}, further to
$O(\log \log \log n)$ by Lee and Liggett~\cite{LeeL13}, and eventually to a
(large) constant by Bilo {\em et al.}~\cite{BiloFM13} (see also Li~\cite{Li09}).

Chekuri~{\em et al.} \cite{ChekuriCLNO07} initiated the line of inquiry
of whether natural game dynamics can lead to efficient
\nash states. 
They considered the following two-phase dynamics: in the first phase, agents arrive in
sequence and choose their best response path upon arrival, and in the second phase, agents
can change their routing path to lower their shared cost (called ``moves'').
Chekuri~{\em et al.} \cite{ChekuriCLNO07} showed that the resulting equilibrium costs at most $O(\sqrt{n}\log^2 n)$ times the cost of the \mst, a factor that was later improved to $O(\log^3 n)$ by Charikar {\em et al.}~\cite{CharikarKMNS08}\footnote{This result applies to a more general setting called
{\em multicast games}, where agents reside at any subset of vertices in the graph.}.
A lower bound of $\Omega(\log n)$ for this ratio was also given by \cite{CharikarKMNS08}, building on a
known lower bound for the online Steiner tree problem~\cite{ImaseW91}.
%
A different approach was taken by Balcan ~{\em et al.} \cite{BalcanBM13}, who considered the problem of influencing
game dynamics in network design games with Shapley sharing, in order to achieve a socially efficient equilibrium.
In their model, players use expert learning, choosing between a best response expert and a central authority expert
suggesting (near-)optimal global behavior. 
At a high level, our results are also for these two distinct approaches -- we show a lower bound for a natural
game dynamics with arbitrary arrivals and departures of agents, and then show an exponentially better upper
bound if the central authority can suggest moves between successive arrival/departure phases.

The analysis of game dynamics in this paper crucially relies on the construction of a hierarchial family of multiple dual solutions.
This method of analysis has been highly influential in designing online algorithms for network design problems.
Implicit use of this method dates back to the work of Imase and Waxman~\cite{ImaseW91} on online Steiner trees,
and a subsequent line of work of \cite{AwerbuchAB04, BermanC97, NaorPS11}.
More recently, this method has been explicitly employed in solving a range of node and
edge-weighted Steiner network design problems in the online setting~\cite{HajiaghayiLP13,HajiaghayiLP14,EneCKP15}.
In terms of the exact techniques, perhaps the closest to our work is that of Umboh~\cite{Umboh15}, who uses
hierarchical tree embeddings to analyze greedy-like algorithms for network design problems. In contrast
to these applications in competitive analysis where decisions are irrevocable, our application in game
dynamics allows temporary overcharging of dual solutions, of which we take advantage in this work.


\section{Model and \ep Dynamics}
\label{sec:model}

In this section and the next we describe and analyze \ep dynamics for
the broadcast game. We first set up our notation and terminology, and
prove some basic structural properties that are used in the rest of
the paper. Let $G=(V,E)$ be a complete graph, $|V|=n$, with metric
costs $c:V\times V\rightarrow \RR_+$ defined on the edges. We assume
without loss of generality that every vertex has a unique agent
(a.k.a. terminal) residing at it. The graph $G$ is revealed via an
online process that is divided into {\em epochs} (indexed by time
$t$). At the start of epoch $t$, the set of vertices in $G$ that have
already appeared is denoted by $V_t$. We denote the set of active
terminals among them by $A_t \subseteq V_t$, i.e., those vertices
whose agents are present in the game at the current epochs.  Each
terminal $v\in A_t$ has a current routing path $p_v$ connecting it to
the common root $r$.  The cost share of $v$ along this routing path is
the sum of $v$'s cost share over the edges in the path, where the cost
of an edge is equally shared between all terminals currently using the
edge. In the \ep scenario, we further enforce the invariant that the
set of paths $p_v$ are in \nash, i.e., no terminal has an incentive to
unilaterally deviate to a different routing path.

The {\em routing} at any time $t$ is defined to be the set of
routing paths $(p_v)_{v\in A_t}$. A {\em best response path} of a terminal $v$ with respect to a routing,
denoted $p^*_v$, is a path from $v$ to $r$ with the minimum shared
cost if $v$ were to move to this path. If there are multiple such
paths, we break ties in favor of paths  having fewer edges with no
terminal other than $v$ using them. Note that this may not break all ties, in which case, any of these paths can be designated
as the best response path.
A terminal $v\in A$ is said to have an {\em improving move} with respect to a routing if by moving
from its current path $p_v$ to a new path $q_v$ strictly decreases
$v$'s cost share.
Given a routing, its potential
\cite{Rosenthal73} is defined to be $\Phi = \sum_{e \in E}
\sum_{i=1}^{N_e} c_e/i$, where $N_e$ is the number of agents using
$e$. A standard argument shows that any improving move decreases the potential by bound that is uniformly bounded away from zero resulting in a finite convergence of our dynamics. 
The following well-known lemma states that in equilibrium the routing
paths always form a tree.

\begin{lemma}
\label{lma:eq-tree}
In equilibrium, the routing paths of a broadcast game form a tree.
\end{lemma}

Each epoch $t$ is divided into several phases. The first phase
consists of an arrival or departure event. In the former case, a new
set of terminals $U_t\subseteq V\setminus V_t$ arrive, and the cost of
all edges incident on terminals in $U_t$ is revealed. Each new
terminal $u\in U_t$ chooses a {\em best response} routing path
$p_u$. In the latter case, a set of terminals leave, thereby removing
the corresponding vertices from the set of terminals $A_t$. (Note that
the corresponding vertices remain in $V_t$.) Lemmas~\ref{lma:leaf} and
\ref{lma:dep+arr} establish that the structure of the set of routing
paths after arrivals or departures remains a tree.

Both arrival and departure events lead to changes in the cost shares
of edges. In the \ep scenario, this might lead to a violation of the
equilibrium state that was being previously maintained.  In this case,
the system performs a sequence of {\em improving moves}, in each of
which a terminal changes its routing path in order to reduce its cost
share.

Improving moves may temporarily create cycles in the collection of
routing paths $\{p_v\}_{v\in A_t}$. We order and group improving moves
into contiguous blocks or phases such that every phase ends with the
routing paths forming a tree. Furthermore, the trees at the beginning
and end of the phase differ in a single pair of edges. The collection
of moves in each such phase is called a {\em tree-follow} move.

\begin{definition}[Tree-follow moves]
  A tree-follow move from $u$ to $v$ in $T$ is a collection of
  improving moves that start with routing tree $T$ and end with
  routing tree $T'=T\setminus (u,\text{parent}(u))\cup (u,v)$, where
  $\text{parent}(u)$ is the parent vertex of $u$ in $T$. Observe that
  each terminal in the subtree rooted at $u$ in $T$ reroutes its path
  to the root in the new rerouting $T'$. (See Figure~\ref{foo1} and
  Figure~\ref{foo2} in Appendix~\ref{sec:tree-move-figs} for an example.)
\end{definition}

A priori, it is not clear whether improving moves can always be grouped
into tree-follow moves. In Lemma~\ref{lma:single-edge}, we show that
in every routing tree $T$ which is not in equilibrium, there exists a
sequence of improving moves that collectively form the tree-follow
move from $u$ to $v$ for some vertices $u$ and $v$. In algorithm \stm,
we use a careful charging scheme to identify the order in which
tree-follow moves should be implemented.

Since every vertex in a tree has a unique path to the root, it
suffices to specify the tree itself in lieu of all of the routing
paths. Henceforth, we will use $T_t$ to denote the tree induced by
$\{p_v\}_{v\in A_t}$ without explicitly specifying the paths
themselves.


\begin{framed}
\textbf{\ep Dynamics}
\begin{enumerate}
\item \textit{Initialization.}  $t=1$, $V_0=\{r\}$, $T_0=\{r\}$, $A_0=\emptyset$.
\item \textit{For $t=1,2,\ldots$}
\begin{itemize}
\item ({\bf Arrivals}.) Let $U_t$ be the set of terminals arriving. Let $A_{t}\gets A_{t-1}\cup U_t$. For each $v\in U_t$, let $p_v=p_v^*$ where $p_v^*$ the best response path with respect to $T_{t-1}$. Let $T_{t}=T_{t-1}\cup_{v\in U_t} p_v^*$. 
\item ({\bf Departures}.) Let $D_t$ be the set of terminals arriving. Let $A_{t}=A_{t}\setminus D_t$. Let $T_{t}=\cup_{v\in A_{t}} p_v$.
\item ({\bf Tree Follow Moves}.)   While $T_t$ is not in equilibrium:\\
  Use algorithm \stm to determine a tree-follow move to implement in
  $T_t$; let this be a move from $u$ to $v$, and let
  $\text{parent}(u)$ denote the parent of $u$ in $T_t$. Implement the
  sequence of improving moves for this tree-follow move to obtain the
  new routing tree $T_t\gets T_t\setminus (u,\text{parent}(u))\cup
  (u,v)$.
\end{itemize}
\end{enumerate}
\end{framed}

Because of departure events, the routing tree may contain non-terminal
vertices as Steiner vertices. It is convenient to extend the notion of
an improving move to vertices that are not terminals. Let $w\notin A$
be a non-terminal vertex. We say that $w$ has an improving move if the
following properties hold: (1) There exists a terminal $v$ whose
routing path $p_v$ includes $w$; let $p_w$ denote the segment of $p_v$
between $w$ and $r$; (2) There exists a path $q_w$ between $w$ and $r$
such that if $v$ were to retain its current routing path from $v$ to
$w$ but move from $p_w$ to $q_w$, then the cost share of $v$ would
strictly decrease.

\subsection{Charging Scheme and Classification of Tree Solutions}
\label{sec:classification}

In proving the upper bound for \ep dynamics, we use a dual
charging scheme to bound the cost of the routing tree. We define the
dual and the corresponding lower bound on the optimal cost next. We
call a partition $P = (S_1,\cdots, S_m)$ of the vertex set $V$
a {\em level-$j$} dual for an integer $j$ if it satisfies the
following properties:
\begin{itemize}
\item $P$ is a partition: $\cup_{S\in P} S = V$, and for any $S_a,
 S_b\in P$, $S_a\cap S_b=\emptyset$.
\item The components have bounded diameter: for any $S\in P$, and any
 vertices $x,y\in S$, $c(x,y)< 2^j$.
\item The components are far from each other: there exists a
 ``center'' $s_i$ in each component $S_i$, such that for all $S_a,
 S_b\in P$, $c(s_a,s_b)\ge 2^{j-1}$.
\end{itemize}
We use the term {\em cuts} to denote the components $S$ of the
partition. The lemma below follows immediately from the observation
that any spanning tree over $V$ must connect the centers of all of the
cuts in a level-$j$ dual $P$.
\begin{lemma}
\label{l:dual-cost}
For any level-$j$ dual $P$, the cost of the minimum spanning
tree \opt is at least $2^{j-1}(|P|-1)$.
\end{lemma}

In order to bound the cost of an equilibrium resulting from \ep, we
relate the cost of the edges used in the solution to a family of
duals. Let $\Pi = \{P_j\}_{j\in\mathbb Z}$ denote a family of
partitions, where $P_j$ is a level-$j$ dual.

Our charging scheme for routing solutions that form a tree proceeds as
follows. Every vertex in the routing tree is responsible for the cost
of its parent edge. Consider an edge $e=(v,\text{parent}(v))$ with
length in $[2^{j+2},2^{j+3})$ for some $j\in\mathbb Z$. We charge the
cost of this edge to the cut in the level-$j$ dual that contains $v$:
$S\in P_j$ such that $v\in S$. Our goal is to show that every cut gets
charged a small number of times.
\begin{lemma}
\label{l:cut-charging}
 Suppose that our charging scheme charges each cut in the family
 $\Pi$ at most once. Then the cost of the solution is at most $O(\log n)$\opt.
\end{lemma}

For much of our analysis, we will assume that the dual family
$\Pi$ is provided to us. In Section~\ref{sec:online} we discuss how to
construct this family algorithmically as terminals arrive online.

\paragraph{Classification of a Tree Routing.}
We classify the tree routings reachable via \ep dynamics into one of
four states depending on the charging structure defined by the
solution. We remark that not all tree routings are reachable via \ep
dynamics, indeed even the set of equilibria obtained is smaller than
the set of all equilibria. Let $T$ be a routing tree for some set of
active terminals $A$. We say a vertex $u$ is a leaf (non-leaf) if it
is a leaf (non-leaf) in $T$. Note that all leaves must be terminals,
but a non-leaf vertex may or may not be a terminal.

\begin{enumerate}
\item \equilibrium: In this state, no terminal (and therefore, no
 non-terminal vertex in $T$) has an improving move.  Furthermore, every cut
 is charged at most once. (Note that not every \nash is a \equilibrium
 state.)
\item \amortized: In this state, some terminals (and potentially
 non-terminals) may have improving moves, but every cut is charged at
 most once.
\item \leaf: In this state, every cut is charged by at most one
 non-leaf vertex (and any number of leaf terminals). (Recall that
 leaf vertices in the routing tree are necessarily terminals.)
\item \nonleaf: In this state, all but one of the cuts are charged by
 at most one non-leaf vertex (and any number of leaf terminals). The
 exceptional cut, that we denote by $S^*$, is charged by at most two
 non-leaf vertices, say $u$ and $v$ (and any number of leaf
 terminals). One of these, $u$ or $v$, must be the last vertex to have made a
 (tree-follow) move.
\end{enumerate}
Note that: \equilibrium$\subseteq$ \amortized$\subseteq$ \leaf$\subseteq$ \nonleaf,
where $A\subseteq B$ implies that a routing tree in state $A$ is also in state $B$.

\subsection{Selecting a Tree-Follow Move}

To define the tree-follow move performed in a non-equilibrium tree state $T$,
we establish a system of priorities among the improving tree moves
based on the current state of the routing tree.
A tree follow move of $u$ to $v$ is said to be a {\em leaf move} if $v$
is a leaf in $T$, and a {\em non-leaf move} otherwise.
%

\begin{framed}
\textbf{Algorithm \stm}
\begin{enumerate}
	\item \equilibrium:\label{move:equil} No terminal has an improving move. The system can
		deviate from an equilibrium state only via arrivals or departure events.
	\item \amortized:\label{move:amort} In this state, for any vertex $u$ that has
         an improving tree move, move $u$ to the closest vertex to
         which it has an improving move.
	\item \leaf:\label{move:int}
		 \begin{enumerate}
		 	\item If there exists a leaf terminal $u$ with a non-leaf move,
		 		then make any such move for $u$.\label{move:int1}
			\item Else, if there exists a non-leaf vertex
                          $u$ with a non-leaf move then move $u$ to
                          the closest such non-leaf $v$. \label{move:int2}
                       \item Else, if there exists a non-leaf
                               vertex $u$ and a leaf terminal $v$
                               such that $u$ and $v$ are charging the
                               same dual cut, then move $u$ to $v$.
                               (Claim~\ref{lem:intermediate1} shows
                               that either $u$ to $v$ or $v$ to $u$
                               is an improving move; The latter is
                               covered by case \eqref{move:int1}.) If
                               there are multiple such leaf terminals
                               $v$, then make any such
                               move.\label{move:int3}
			\item Else, make any improving move. (This
                         will necessarily be a leaf-to-leaf move by
                         exclusion of the previous three cases.)\label{move:int4}
		\end{enumerate}
       \item \nonleaf: Let $u$ and $v$ be the non-leaf vertices
               that are charging the special cut $S^*$. If $u$ has an
               improving move to $v$ then move $u$, else move $v$,
               in either case to the closest vertex to which they
               have an improving move. (We show in
               Claim~\ref{lemma:two-non-leaf} that one of $u$ and $v$
               has an improving move.)\label{move:nonleaf}
\end{enumerate}

\end{framed}

The validity of the algorithm depends on the following two claims that
we prove in Appendix~\ref{sec:defer}. The first claim shows that whenever a cut is being
charged by a leaf and a non-leaf, at least one of these two vertices
has an improving move to the other. In this case, we can find a valid
tree-move for Step~\eqref{move:int3} of \stm. The second claim shows
that in a \nonleaf state, whenever a cut is being charged by two
non-leaves, at least one of these two vertices has an improving move
to the other; we can then find a valid tree-move for
Step~\eqref{move:nonleaf} of \stm.

\begin{claim}\label{lem:intermediate1}
Let the routing tree $T$ be in \nonleaf state but not in \amortized state. 
Let $u, v$ be a pair of vertices in $T$ charging the same dual cut $D$ at
some level $i$, where at most one of $u$ or $v$ is a non-leaf vertex in $T$.
Then, at least one of $u$ moving to $v$ and $v$ moving to $u$ is an improving move.
\end{claim}

\begin{claim}\label{lemma:two-non-leaf}
  Let $T$ be a routing tree in a \nonleaf but not a \leaf state,
  arrived at by implementing one of the steps
  \eqref{move:amort}--\eqref{move:nonleaf} in algorithm \stm.  Let $u$ and $v$ be the
  non-leaf vertices charging the special cut $S^*$. Then either $u$ to
  $v$ or $v$ to $u$ is an improving move.
\end{claim}

\section{Analysis of the \ep Dynamics}
\label{sec:ep-analysis}

In this section we prove Theorem~\ref{thm:ep}. Some proofs from this
section are deferred to Appendix~\ref{sec:defer}.

Our argument hinges on a closure property: the epoch starts with the
routing tree being in the \equilibrium state;
Lemma~\ref{lma:single-edge} argues that whenever the current routing
tree is not in equilibrium, at least one improving move exists, and we
can use algorithm \stm to make a move; Lemma~\ref{lem:closure} then
shows that for the moves made by algorithm \stm, the routing tree
remains in one of the four states defined above, in particular, it is
always in a \nonleaf state. 
The epoch ends when the routing tree re-enters a \equilibrium
state. At this point, by definition, each dual cut is charged at most
once, and therefore, by Lemma~\ref{l:cut-charging} the cost of the
routing tree is bounded, and Theorem~\ref{thm:ep} follows. We must
also argue termination of the sequence of moves, but this follows
directly from a standard potential argument based on the fact that all
our moves are improving moves. 


\paragraph{The arrival or departure phase.}
We first consider the situation where the epoch begins with an arrival
event. Every arriving terminal $u$ chooses its best response path
$p^*_u$ as its current routing path $p_u$. We claim that the new
routing paths $\{p_u\}_{u\in U_t}$ along with the current routing tree
$T$ continue to form a tree solution and we end up in a \leaf state.

\begin{lemma}
\label{lma:leaf}
	Suppose a set of new terminals $U_t$ arrive in epoch $t$ when the routing paths of the
	existing terminals are in an equilibrium state. Then, for each new terminal $u\in U_t$,
	the chosen routing path $p_u$ comprises a single edge $(u, v)$
	connecting $u$ to an existing vertex $v$ in the current routing tree $T$, and then
	following the unique path in $T$ from $v$ to $r$.
\end{lemma}


\noindent
We therefore obtain the following lemma.

\begin{lemma}
\label{lma:dep+arr}
  After the arrival or departure of a set of terminals in an
  \equilibrium state, the routing tree $T$ remains in a \leaf state.
\end{lemma}

\paragraph{Sequence of tree-follow moves.} We now consider improving
moves made by the algorithm \stm. We first observe that for any
routing tree that is not at equilibrium, there must exist an improving
tree-follow move; see Appendix~\ref{sec:defer} for a proof. 


\begin{lemma}
\label{lma:single-edge}
	If the routing tree is not in equilibrium, then at least one improving
	tree-follow move exists.
\end{lemma}



\noindent
We therefore have the following simple observation.
\begin{observation}
\label{lma:tree}
	In \ep dynamics the routing paths at the end of a phase always form a tree.
\end{observation}

\noindent
We are now ready to establish our main technical result of this
section, namely that the four states defined in
Section~\ref{sec:classification} are closed under \ep dynamics.

\begin{lemma}\label{lem:closure}
Let $T$ be the routing tree for which we make an improving tree-move
in Step~\eqref{move:int} of algorithm \ep.
\begin{enumerate}[(i)]
\label{lem:independent} \item If $T$ is in a \amortized state but not
in a \equilibrium state, then after the move selected  in
Step~\eqref{move:amort} of \stm, the resulting routing tree is in a \nonleaf state.
\label{lma:leaf1} \item
	If $T$ is in a \leaf state, then after the move selected in
       Step~\eqref{move:int} of \stm, the resulting routing tree is in a \nonleaf state.
\label{l:nonleaf-move}
\item	If $T$ is in a \nonleaf state,  then after the move selected
 in Step~\eqref{move:nonleaf} of \stm,
	 the resulting routing tree is in a \nonleaf state.
\end{enumerate}
\end{lemma}

\begin{proof}
We complete the proof by a detailed case analysis.
\begin{enumerate}[(i)]
\item Let $T$ be in a \amortized state but not a
  \equilibrium. Therefore, every dual is being charged at most once in
  $T$. After the move, the new tree $T'$ contains exactly one edge not
  in $T$. The charging for this edge can introduce at most one dual
  cut which is charged more than once.
Thus the new routing tree is in a \nonleaf state.

\item Let $T$ be a tree in a \leaf state. We now consider different cases depending on the move chosen by algorithm \stm.
\begin{enumerate}[(a)]
\item[{\bf  Step~\eqref{move:int1}:}] Suppose the algorithm makes a
  leaf to non-leaf move in Step~\eqref{move:int1}, then the only new edge introduced
is $u$'s parent edge. Since $u$ remains a leaf in $T$ and its new
parent was already a non-leaf vertex, no new
non-leaf vertex is introduced. Thus, in the new routing tree, every dual is charged by at most one non-leaf vertex.
This implies that the \leaf state is preserved.
\item[{\bf Step~\eqref{move:int2}:}] Suppose there is no improving
  move in Step~\eqref{move:int1}, and algorithm makes a non-leaf to
  non-leaf move in Step~\eqref{move:int2} for vertex $u$. Let the new
  edge introduced be $(u,v)$ where $v$ is a non-leaf vertex.  After
  this move, the only cut that has an additional non-leaf vertex
  charging to it is the cut being charged by $u$ (call it $S$). Prior
  to this move, $S$ had at most one non-leaf vertex charging it. After
  $u$'s move, it has at most two non-leaves charging it, with one of
  them ($u$) having made the last move. Therefore, $T$ is in a \leaf
  or \nonleaf state.
\item[{\bf Step~\eqref{move:int3}:}] Suppose there are no improving
moves in Steps~\eqref{move:int1} and \eqref{move:int2}, and the
algorithm performs a non-leaf to leaf move in Step~\eqref{move:int3} from $u$ to
$v$. Prior to the move, $u$ and $v$ were charging the same cut, say
$S$ at level $i$. After $u$'s move, $u$'s parent edge, $(u,v)$, is of
length $<2^i$, and therefore, $u$ charges a cut different from $S$
(call it $S'$), whereas $v$ continues to charge $S$.



Therefore, the only cuts that get charged by new non-leaves after
$u$'s move are $S$ and $S'$. $S$ was previously being charged by a single
non-leaf, namely $u$; Now it is charged by only one non-leaf, namely
$v$. $S'$ was previously being charged by at most one non-leaf (by
virtue of the routing tree being in a \leaf state), so now it
is being charged by at most two non-leaves, one of which is
$u$. Therefore, $T$ is in a \leaf or \nonleaf state.


\item[{\bf Step~\eqref{move:int4}:}] Finally, consider the scenario
  where there are no improving moves in any of
  Steps~\eqref{move:int1}, \eqref{move:int2}, and
  \eqref{move:int3}. In this case, the algorithm makes a leaf to leaf
  move in Step~\eqref{move:int4} from $u$ to $v$.  The only new
  non-leaf vertex created by the move is $v$. We first argue that the
  dual charged by $v$ (say $S$) does not have a second non-leaf vertex
  charging it. Since Step~\eqref{move:int3} was not executed, it
  follows that no cut was being charged by both a non-leaf and a leaf
  vertex before the move (although multiple leaf terminals might be
  charging the same cut). In particular, $v$ was a leaf vertex
  charging $S$ before the move, and so, no other non-leaf vertex was
  charging $S$ before or after the move.

  The only other cut that gets a new charge after the move is the cut
  charged by $u$'s new edge. Since $u$ is a leaf, this cut continues
  to have at most one non-leaf charging it. Therefore, the routing
  tree is in a \leaf state.


%
%
%

%
\end{enumerate}

\item Let $T$ be in a \nonleaf state. Recall from the
  definition of Step~\eqref{move:nonleaf} that $u$ and $v$ are the two
  non-leaf vertices charging the special cut $S^*$ at some level $i$,
  and $u$ has an improving move to $v$. Then we have $c_{uv} <
  2^i$. Suppose that $u$ made the improving move to $w$, where $w$ can
  be $v$. Since $u$ chooses the improving move to the closest vertex,
  we have $c_{uw}\leq c_{uv}<2^i$. After the move, $u$ must charge a
  cut, say $S$ whose level is strictly less than $i$. Thus $S\neq
  S^*$, and $S^*$ now has only a single non-leaf charging
  it. Moreover, before the move $S$ had at most one non-leaf charging
  it. Now, along with $u$, it can have two non-leaves charging it but
  one of them, $u$, has made the last move. Thus, the new tree is in
  a \nonleaf state.\qedhere
\end{enumerate}
\end{proof}




\bibliographystyle{plain}

\bibliography{ref}
\appendix
\section{Lower Bound for \nep Dynamics}
\label{sec:nep}

\newcommand{\Final}{{\mathcal F}\xspace}

In this section, we will show that if arrivals and departures are
allowed at non equilibrium states, then no dynamics can lead to a good
equilibrium.

\begingroup
\def\thetheorem{\ref{thm:nep}}
\begin{theorem} {\bf (Restatement)}
  There exists a graph on $n$ vertices with a sequence $\sigma$ of
  $\N=\Theta(n^2)$ arrivals and departures such that if every terminal
  chooses its best path at its time of arrival, then the state of the
  network after the final arrival is an equilibrium and its cost is
  $\Omega(\N^{1/6})$ times the cost of the minimum spanning tree over
  all vertices of the underlying graph.
\end{theorem}
\addtocounter{theorem}{-1}
\endgroup

The rest of this section is devoted to proving Theorem
\ref{thm:nep}. In the previous sections, it was convenient to
assume that players arrive at distinct vertices. For this section, we
will assume that multiple players can arrive at the same vertex. Recall that multiple players at a vertex can be modeled in the single player case using zero cost edges; hence, there is no technical distinction between the two cases.

\begin{figure}[h]
  \centering
  \includegraphics[scale=0.4]{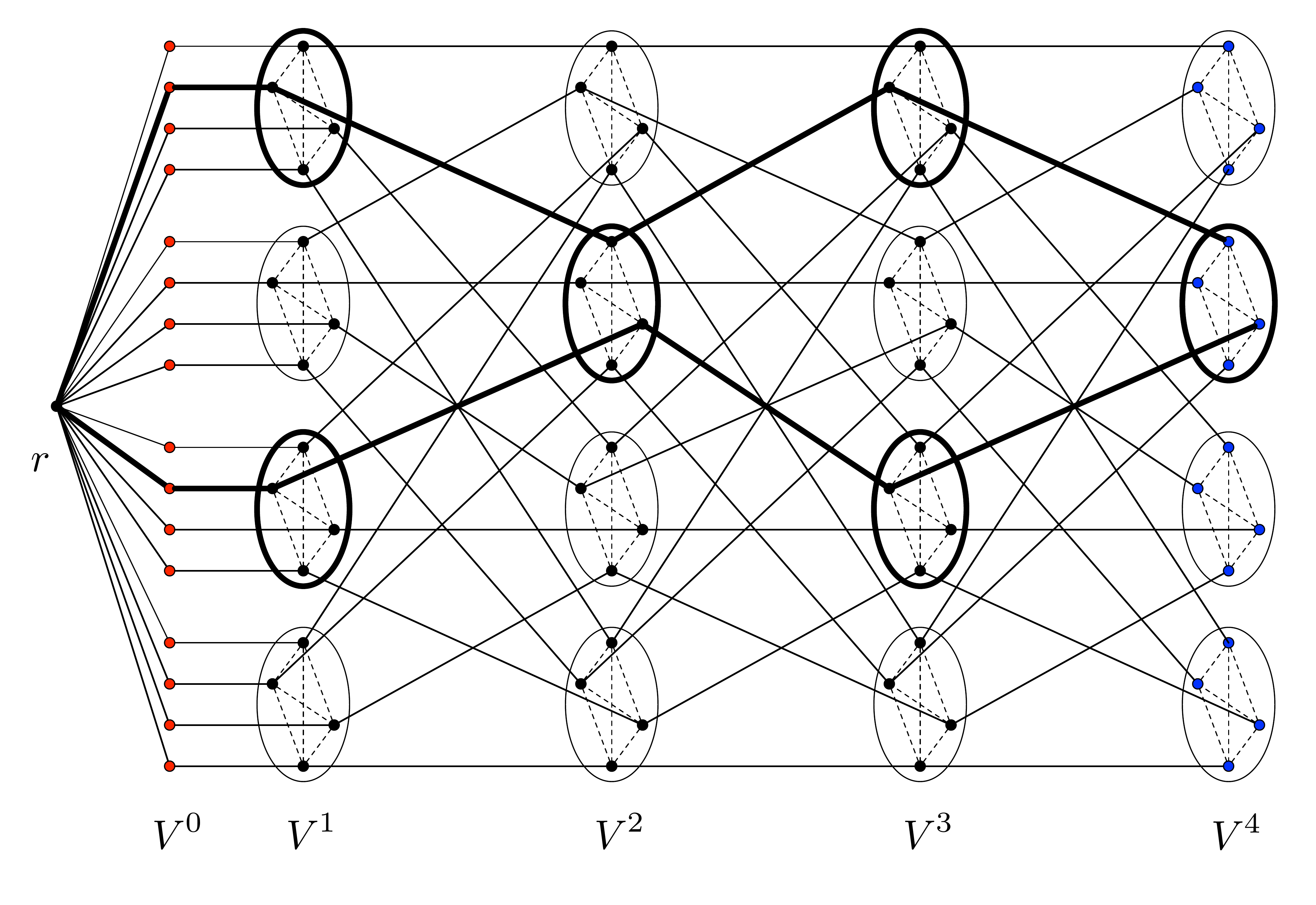}
  \caption{Example for $m = 4$. Auxiliary vertices are in
    red, end vertices are in blue. Ovals represent clusters. Intra-cluster edges are shown as dashed edges. The two bold paths starting from the
    same cluster successively diverge into different clusters and
    converge into the same cluster on their way to the root.}
  \label{fig:lower-bound-zigzag}
\end{figure}

We construct a family of lower bound instances parameterized by an
integer $m\ge 1$. The $m$th instance uses the metric induced by
weighted graph $G_m$ (see Fig~\ref{fig:lower-bound-zigzag}). The vertex set
of this graph consists of a root
$r$ and $m+1$ \emph{layers} $V^0, \ldots, V^m$. For $1 \leq i \leq m$,
layer $V^i$ consists of $m$ \emph{clusters} $C^i_1, \ldots, C^i_m$,
each of which is a clique over $m$ vertices. We use $v^i_{j,k}$ to
denote the $k$-th vertex of $C^i_j$; recall that each of $i$, $j$, and
$k$ take on integral values in $[m]$. Layer $V^0$ also consists of $m^2$
vertices, which are labeled $v^0_{j,k}$ for $j,k \in [m]$, but there
are no edges between these vertices. The
vertices of $V^m$ are called \emph{end} vertices, and those of $V^0$
are called \emph{auxiliary} vertices. Observe that the graph $G_m$ has
$n = m^2(m+1) + 1$ vertices in all.

Next, we describe the edges. Each pair of vertices within the same
cluster $C^i_j$ is connected by an edge of length $1/m$ for all layers except $V^0$. The remaining
edges in the graph connect vertices in neighboring layers and are all
of length $1$. Each auxiliary vertex $v^0_{j,k}$ in $V_0$ is connected
to the root and to its corresponding vertex $v^1_{j,k}$ in layer
$1$. For $1 \leq i \leq m-1$, we have an edge
$(v^i_{j,k}, v^{i+1}_{k,j})$ for
each $j,k \in [m]$. In other words, the vertices of the $j$-th cluster in layer $i$ are connected to the $k$-th vertices of the clusters in layer $i+1$; in particular, the $k$-th vertex of the $j$-th cluster in layer $i$ is connected to the $j$-th vertex of the $k$-th cluster in layer $i+1$. For example, see the edges leaving the first (top) cluster of $V_1$ in Figure \ref{fig:lower-bound-zigzag}. Observe that there are exactly $m^2(m+1)$
inter-layer edges, and exactly $m^3(m-1)/2$ intra-cluster edges.


Observe that each end vertex $v^m_{j,k}$ has a unique path $P_{j,k}$
to the root that consists of only inter-layer edges (see
Figure~\ref{fig:lower-bound-zigzag}). We call these paths {\em canonical
  paths}. Note that each inter-layer edge belongs to exactly one
canonical path. In other words, the set of inter-layer edges is a
disjoint union of all the canonical paths $P_{j,k}$.

\eat{
\begin{figure}[h]
  \centering
  \includegraphics[scale=0.4]{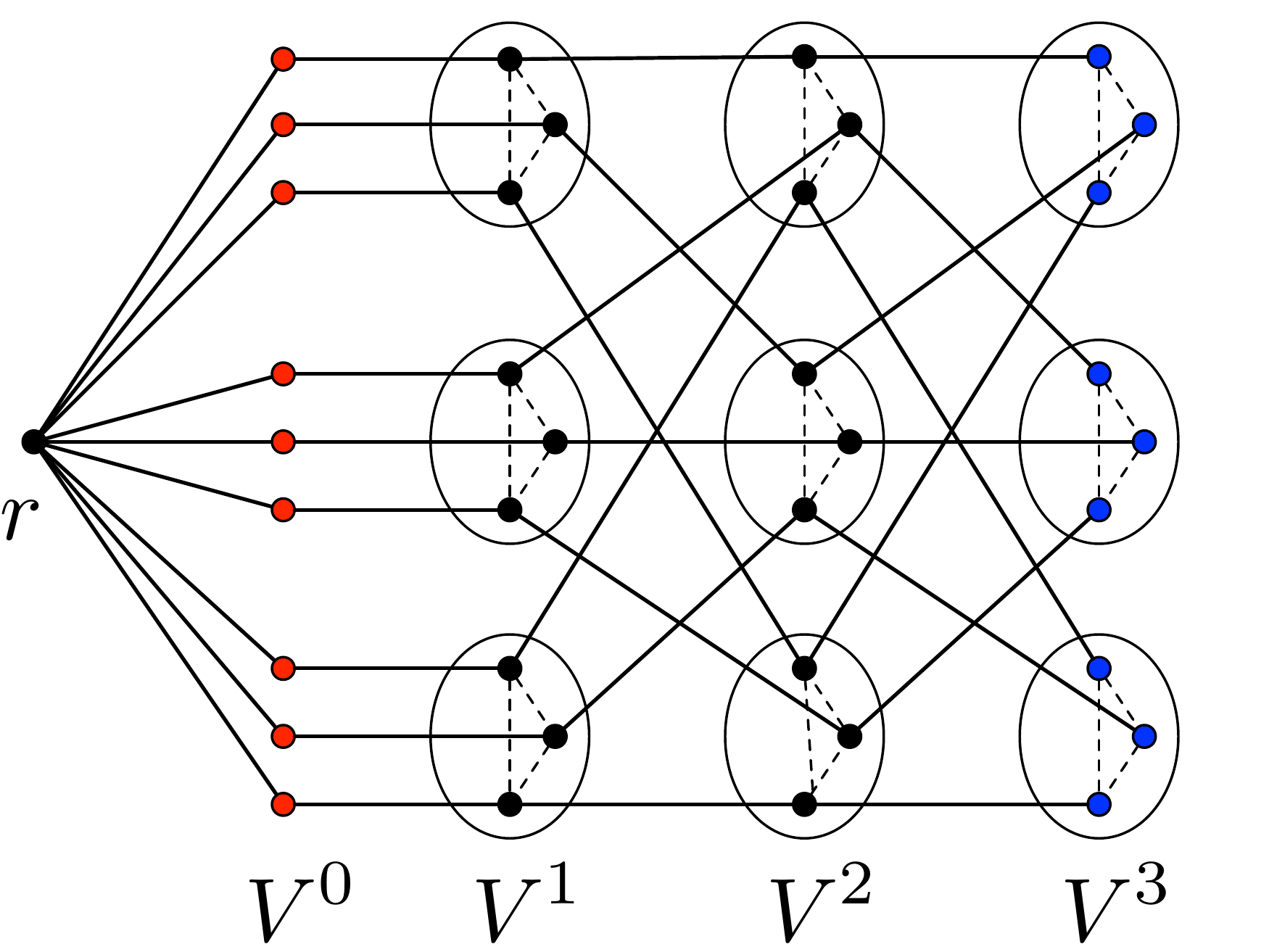}
  \caption{Example for $m = 3$. Auxiliary vertices are in
    red, end vertices are in blue. Ovals represent clusters. Intra-cluster edges are shown as dashed edges.}
  \label{fig:lower-bound}
\end{figure}
}

\eat{\begin{figure}[h]
  \centering
  \includegraphics[scale=0.4]{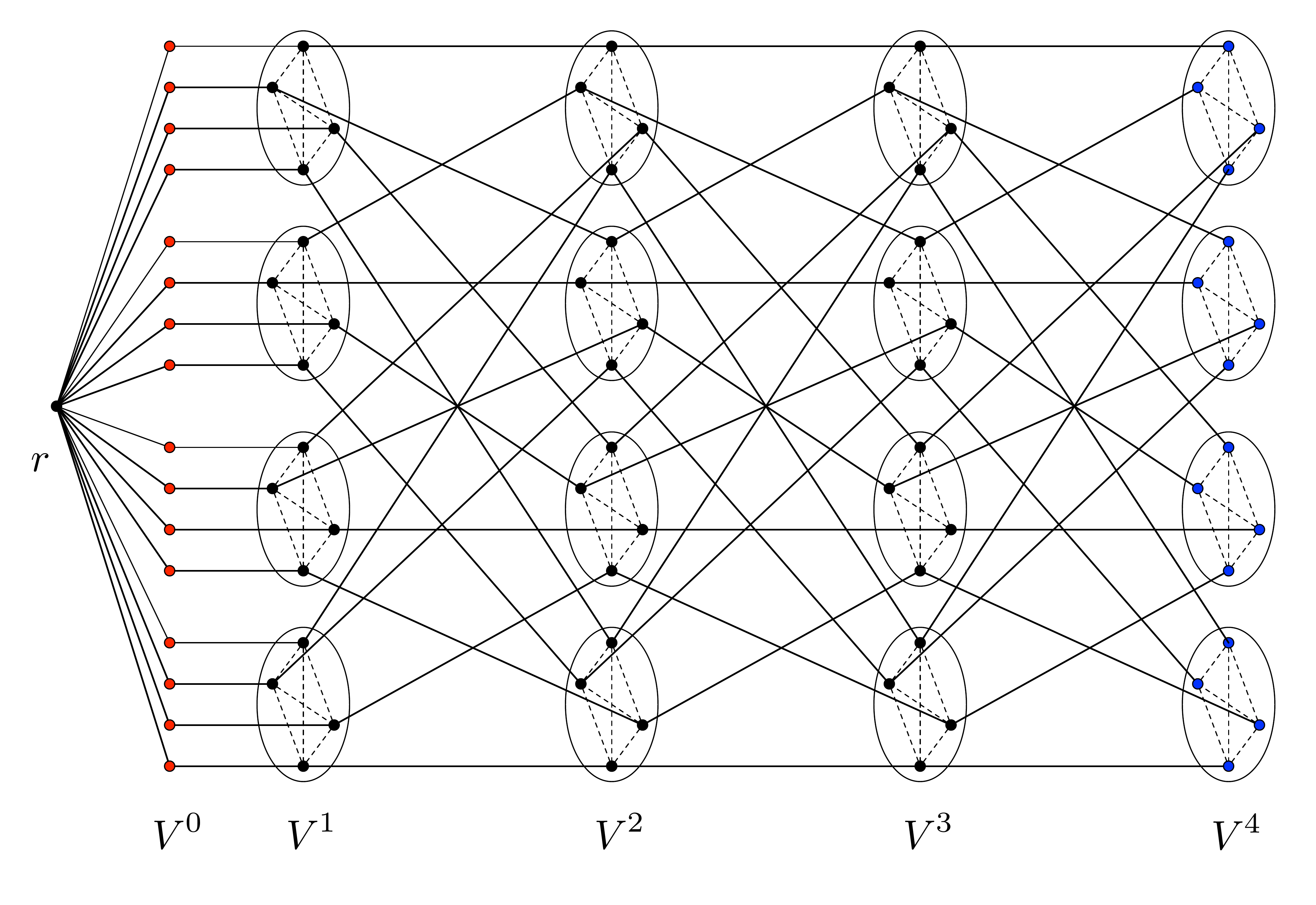}
  \caption{Example for $m = 4$. Auxiliary vertices are in
    red, end vertices are in blue. Ovals represent clusters. Intra-cluster edges are shown as dashed edges.}
  \label{fig:lower-bound}
\end{figure}
}

\paragraph{The cost of the final equilibrium.} Our lower bound
instance consists of a sequence of arrivals and departures of
terminals. Each terminal/player, upon arrival, chooses its best
response path to the root at the time of its arrival. Terminals are
not allowed to change their chosen path until after the sequence
ends. We show that: (1) each terminal chooses its canonical path as
its best response path to the root; (2) at the end of the sequence,
these paths collectively form an equilibrium, and therefore, no
terminal wants to move; (3) this equilibrium state is far from optimal
in cost.

Before we describe the sequence of arrivals and departures in full
detail, we will analyze the final equilibrium state and its cost
relative to the optimal cost. Let $\OPT$ denote the cost of the
minimum spanning tree over all vertices in $G_m$. Observe that this is
an upper bound on the cost of any optimal solution at the end. The
final state following our sequence of arrivals and departures, denoted
$\Final$, consists of $m$ players situated at every end vertex
$v^m_{j,k}$ in layer $m$; each player uses the canonical path
$P_{j,k}$ to route to the root. The following lemma shows that this
is an equilibrium state with a polynomially larger cost relative to
$\OPT$.

\begin{lemma}
  \label{lem:lb-equil}
  State $\Final$ is an equilibrium and the cost of $\Final$ is $\Omega(m)\OPT$.
\end{lemma}
\begin{proof}
  First, we prove that $\Final$ is an equilibrium. Consider a player
  at end vertex $v^m_{j,k}$ with path $P_{j,k}$ and an alternative
  path $p'$. For every intra-cluster edge $e$, we have $N_e(\Final) = 0$, and
  for every inter-layer edge $e$, we have $N_e(\Final) = m$, where
  $N_e$ denotes the number of terminals using edge $e$. So the
  player's current cost share is $\frac{m+1}{m}$.
  Path $p'$ contains at least
  one intra-cluster edge and at least $m+1$ inter-layer edges. Thus, the
  player's cost share when it switches to $p'$ is at least
  $\frac{1}{m} + \frac{m+1}{m+1} = \frac{m+1}{m}$.
  Therefore, $\Final$ is an equilibrium state.

  Next, we prove that $c(\Final) = \Omega(m)\OPT$. $\Final$ consists
  of all unit-length inter-layer edges, and therefore its cost is
  $m^2(m+1)$. On the other hand, one way of constructing a spanning
  tree for $G_m$ is to select an arbitrary spanning tree within each
  of the $m^2$ cliques, all of the edges from layer $0$ to the root,
  as well as one inter-layer edge per clique connecting it to its
  preceding layer, say the edge $(v^i_{j,j}, v^{i+1}_{j,j})$ to
  connect $C^{i+1}_j$ to $C^i_j$ for all $i, j\in [m]$. The total cost
  of this solution is
$$
\frac 1m m^3 + m^2 + m^2 = 3 m^2.
$$
This concludes the proof.
\end{proof}


\paragraph{Sequence of arrivals and departures.}
The sequence is constructed in $m$ \emph{phases}, each phase
consisting of $m^2$ \emph{rounds}, one per end vertex $v^m_{j,k}$, and
indexed by $(j,k)$. Informally, the objective of each phase is to add
one more terminal at each of the end vertices $v^m_{j,k}$. Within
round $(j,k)$ in a phase, we use a set of ``temporary'' terminals whose sole aim is to force the terminal at $v^m_{j,k}$ that arrives at the end
of the round to choose the canonical path as its best response. The
temporary terminals are introduced at intermediate vertices along the
canonical path during the round, and removed at the end of the round.

Formally, let $\prec$ be an arbitrary total order on the pairs
$(j,k)$. The sequence $\sigma$ is constructed to maintain the
following invariant: at the end of round $(j,k)$ of phase $\ell$,
there will be $\ell$ players on $v^m_{j',k'}$ for $(j', k') \prec (j,
k)$, and $\ell-1$ players on the remaining end vertices. Furthermore,
each player on $v^m_{j,k}$ uses the path $P_{j,k}$.

We now specify the subsequence for each round. Consider round $(j,k)$
of phase $\ell$. For simplicity of notation, we use $v^i$ to denote
the vertex of $V^i$ on $P_{j,k}$. We also use $P^i$ to denote the
segment of $P_{j,k}$ starting at $v^i$ and ending at the root. The
round consists of $m+1$ {\em iterations}. In iteration
$0 \leq i\leq m-1$, $m^2$ players arrive at $v^i$. In iteration
$i = m$, one player arrives at $v^m$. Finally, the players on
$v^0, \ldots, v^{m-1}$ depart.

Using induction over the terminal arrivals, we now show that for every
terminal, the best-response path on arrival is the segment of the
canonical path connecting it to the root.

\begin{lemma}
  \label{lem:lb-inv}
  Consider a terminal arriving at vertex $v^i$ in iteration $i$
  of round $(j,k)$ in phase $\ell$. The best-response path of the
  terminal to the root is the segment of its canonical path $P^i$.
\end{lemma}

\begin{proof}
  We prove the invariant by induction over the sequence of terminal
  arrivals, ie, over $(\ell,j,k, i)$. Moreover, we will only prove
  the that the inductive hypothesis holds for the first terminal
  arriving in an iteration; every subsequent terminal will clearly
  choose the same path as the first terminal since they are arriving
  at the same vertex. The invariant trivially holds prior to the start of
  round $1$ of phase $1$.

  Now consider the start of iteration $i$ of round $(j,k)$ in phase $\ell$ and assume
  the invariant held in previous iterations, rounds and phases. Let us
  count the number of terminals on each edge. So far, on each end
  vertex, there are either $\ell$ or $\ell-1$ terminals and each of
  them chose their canonical path on arrival, by the inductive
  hypothesis. The inductive hypothesis also tells us that for each
  $i' < i$, there are at
  least $m^2$ terminals on $v^{i'}$ using the path
  $P^{i'}$. Thus, each inter-layer edge belonging to $P^{i-1}$ has at least
  $m^2$ terminals, and each inter-layer edge that does not belong to $P^{i-1}$ has at
  most $\ell$ terminals. Moreover, none of the intra-cluster edges are used by any
  terminal.

\eat{Since
  terminals only reside at end vertices
  for all canonical paths except $P_{j, k}$, and each of these terminals
  chose their canonical path on arrival (by the inductive hypothesis),
  every edge that is not in $P^{i-1}$ has either $\ell$ or $\ell-1$
  terminals on it.
  Furthermore, every edge in $P^{i-1}$ has at least
  $m^2$ terminals on it.
}

  Note that $P^i$ consists of the inter-layer edge $(v^i, v^{i-1})$ followed by
  $P^{i-1}$. Since
  $|P^{i-1}| \leq m$, the cost share of the new terminal at $v^i$
  (call this terminal $a$) on $P^i$ is at most
  $1/\ell + m/(m^2 + 1) < 1/\ell + 1/m$. Any other path $Q$ for $a$
  contains at least two inter-layer
    edges that do not belong to $P^{i-1}$ and at least one intra-cluster
    edge, so $a$'s cost share on $Q$ is at least
    $2/(\ell + 1) + 1/m \ge 1/\ell + 1/m$.
    Thus, player $a$'s unique best-response path is $P^i$.
\end{proof}

Lemma \ref{lem:lb-inv} shows that the sequence of arrivals and
departures described above terminates in the final state $\Final$,
which costs $\Omega(m)\OPT$ by Lemma \ref{lem:lb-equil}. This
completes the proof of Theorem \ref{thm:nep}.

\section{Deferred Proofs}
\label{sec:defer}
We will now present proofs that were skipped in Sections~\ref{sec:model} and \ref{sec:ep-analysis}.

\begingroup
\def\thetheorem{\ref{lma:eq-tree}}
\begin{lemma}
In equilibrium, the routing paths of a broadcast game form a tree.
\end{lemma}
\addtocounter{theorem}{-1}
\endgroup
\begin{proof}
	The lemma is a direct consequence of the following downward closure property that
	holds in an equilibrium state. Suppose $w$ is vertex (not necessarily a terminal)
	that appears on the routing paths $p_u$ and $p_v$ of two terminals $u$ and $v$ respectively.
	Then, the segment of $p_u$ and $p_v$ between $w$ and $r$ must be identical.
	For the sake of contradiction, suppose this claim is false, and let $q_u$ and $q_v$
	denote the non-identical segments of $p_u$ and $p_v$ between $w$ and $r$. Assume
	wlog that the current shared cost of $q_u$ is at most that of $q_v$. If $v$ now moves
	to a new routing path that follows $p_v$ until $w$ and then uses $q_u$ to reach
	$r$, then the change in cost of $v$ will be the difference of the new shared cost of
	$q_u$ and the current shared cost of $q_v$. This is clearly non-positive by our
	assumption on the relative order of current shared costs of $q_u$ and $q_v$.
	In fact, we argue that this difference is negative. First, note that there is
	at least one edge that is in $q_u$ but not in $q_v$ since the two paths are
	non-identical. Now, let $(x, y)$ be the closest edge to $u$ in $q_u$ that does not
	appear in $q_v$, where $x$ is closer to $u$
	than $y$. Then, $x$ must also appear on $q_v$, and therefore, cannot appear
	in the segment of $p_v$ between $v$ and $w$. It follows that edge $(x, y)$
	does not appear in the segment between $v$ and $w$ in $p_v$, and hence is absent
	from the entire path $p_v$. When $v$ moves to its new path, the shared cost on
	$(x, y)$ {\em decreases} below its current value, and therefore, the shared
	cost of $q_u$ decreases as a whole. This implies that $v$ has an improving move,
	which contradicts the premise that the terminals are in an equilibrium state.
\end{proof}

\begingroup
\def\thetheorem{\ref{l:cut-charging}}
\begin{lemma}
Suppose that our charging scheme charges each cut in the family
 $\Pi$ at most once. Then the cost of the solution is at most $O(\log n)$\opt.
\end{lemma}
\addtocounter{theorem}{-1}
\endgroup
\begin{proof}
 Let $D$ denote the largest edge length in the graph. Then, we first
 note that we may ignore edges in our solution of length at most
 $D/n$. This is because there are at most $n$ such edges, and \opt
 is at least $D$ by the metric property of edge lengths. For the
 remainder, we charge duals at levels $j$ for $j\in (\log(D/n)-3,
 \log D-2]$. There are at most $\log n+1$ such duals. The total cost of
 edges charged to a dual at level-$j$ is at most
 $2^{j+3}|P|<32\,2^{j-1}(|P|-1)$. Lemma~\ref{l:dual-cost} then implies the result.
\end{proof}

\begingroup
\def\thetheorem{\ref{lem:intermediate1}}
\begin{claim}
Let the routing tree $T$ be in \nonleaf state but not in \amortized state.
Let $u, v$ be a pair of vertices in $T$ charging the same dual cut $D$ at
some level $i$, where at most one of $u$ or $v$ is a non-leaf vertex in $T$.
Then, at least one of $u$ moving to $v$ and $v$ moving to $u$ is an improving move.
\end{claim}
\addtocounter{theorem}{-1}
\endgroup
\begin{proof}
Let $p_u$ and $p_v$ be the routing paths of
$u$ and $v$ respectively, where $u$'s parent edge is $(u,x)$ and $v$'s parent edge
is $(v,y)$. Then, we have $c_{ux}, c_{vy}\in
[2^{i+2},2^{i+3})$. Moreover, since $u$ and $v$ belong to the same cut
at level $i$, $c_{uv} < 2^{i}$.
Therefore,
\begin{equation}
\label{eq:cont2}
	c_{ux}, c_{vy} > 4 c_{uv}.
\end{equation}
Let $N_e$ be the number of vertices using edge $e$.
Since at most one of $u$ or $v$ is a non-leaf in $T$, we can assume wlog that
$u$ is a leaf in $T$ (otherwise, the contradiction that we derive below for edge
$(u, x)$ can be derived instead for edge $(v, y)$). In other words, $N_{ux} = 1$.

For the sake of contradiction, let us assume that neither $u$ moving to $v$, nor $v$
moving to $u$ is an improving move. Since $u$ moving to $v$ is not an improving move, we have the shared cost on path $p_u$ is at most the shared cost for $u$ if it shifts to the path $(u,v)\cup p_v$. Formally, we have
\begin{eqnarray}
	c_{ux}+ \sum_{e\in p_u\setminus (u,x)} \frac{c_e}{N_e}
	& \leq & c_{uv} + \sum_{e\in p_v\setminus p_u} \frac{c_e}{N_e+1} + \sum_{e\in p_v\cap p_u} \frac{c_e}{N_e}\nonumber\\
	\text{i.e., } \quad c_{ux}+ \sum_{e\in (p_u\setminus  (u,x))\setminus p_v} \frac{c_e}{N_e} + \sum_{e\in p_u\cap p_v} \frac{c_e}{N_e}
	& \leq & c_{uv} + \sum_{e\in p_v\setminus p_u} \frac{c_e}{N_e+1} + \sum_{e\in p_v\cap p_u} \frac{c_e}{N_e}\nonumber\\
	\text{i.e., } \quad c_{ux}+ \sum_{e\in (p_u\setminus  (u,x))\setminus p_v} \frac{c_e}{N_e}
	& \leq & c_{uv} + \sum_{e\in p_v\setminus p_u} \frac{c_e}{N_e+1}.\label{eq:leaf1}
\end{eqnarray}
Similarly, since $v$ moving to $u$ is not an improving move, we have
\begin{eqnarray}
	\sum_{e\in p_v} \frac{c_{e}}{N_e}
	& \leq & c_{uv} + \frac{c_{ux}}{N_{ux} + 1}+ \sum_{e\in (p_u\setminus (u,x))\setminus p_v} \frac{c_e}{N_e+1} + \sum_{e\in p_u\cap p_v} \frac{c_e}{N_e}\nonumber\\
	\text{i.e., } \quad \sum_{e\in p_v\setminus p_u} \frac{c_{e}}{N_e} + \sum_{e\in p_v\cap p_u} \frac{c_{e}}{N_e}
	& \leq & c_{uv} + \frac{c_{ux}}{N_{ux} + 1}+ \sum_{e\in (p_u\setminus (u,x))\setminus p_v} \frac{c_e}{N_e+1} + \sum_{e\in p_u\cap p_v} \frac{c_e}{N_e}\nonumber\\
	\text{i.e., } \quad \sum_{e\in p_v\setminus p_u} \frac{c_{e}}{N_e}
	& \leq & c_{uv} + \frac{c_{ux}}{N_{ux} + 1}+ \sum_{e\in (p_u\setminus (u,x))\setminus p_v} \frac{c_e}{N_e+1}.\label{eq:leaf2}
\end{eqnarray}
Adding inequalities \eqref{eq:leaf1} and \eqref{eq:leaf2}, and replacing $N_{ux} = 1$, we get
\begin{eqnarray*}
	c_{ux}+ \sum_{e\in (p_u\setminus (u,x))\setminus p_v} \frac{c_e}{N_e} + \sum_{e\in p_v\setminus p_u} \frac{c_{e}}{N_e}
	& \leq & c_{uv} + \sum_{e\in p_v\setminus p_u} \frac{c_{e}}{N_e+1} + c_{uv} +  \frac{c_{ux}}{2}+ \sum_{e\in (p_u\setminus (u,x))\setminus p_v} \frac{c_e}{N_e+1} \\
	\text{i.e., } \quad c_{ux} & \leq & 4 c_{uv},
\end{eqnarray*}
which contradicts inequality~\eqref{eq:cont2}.
Thus, it must be the case that at least one of $u$ moving to $v$
or $v$ moving to $u$ is an improving move.
\end{proof}

We will prove Claim~\ref{lemma:two-non-leaf} next. Before we restate
and prove that claim, we show a stability property of the improving
moves made by algorithm \stm, namely that moving a terminal $u$ does
not create any new improving moves for it.

\begin{claim}
\label{lma:tree-follow}
	Suppose that in the current routing tree, vertex $u$ moving to vertex
	$v$ is {\bf not} an improving move for $u$. Then, after $u$ moves to
	some other vertex $x$ (using a tree-follow move), $u$ moving to $v$
	is still not an improving move.
\end{claim}	
\begin{proof}
	Let $p_u$ and $p_v$ denote the paths connecting $u$ and $v$ respectively to $r$
	in the routing tree $T$ after $u$'s move. Let $N_e$ and $N'_e$ be the number of
	terminals, except $u$, that are routing through edge $e$ before and after $u$'s
	tree move respectively.
	For all edges in $p_u$, we have $N'_e \geq N_e$; for all other edges in $T$,
	$N'_e \leq N_e$. Since $u$ moving to $v$ was not an improving move but
	moving to $x$ was, the shared cost of $u$ if it moved to $v$ would have been more
	than its shared cost after the tree move to $p_u$. Hence,
	\begin{eqnarray*}
		c_{uv} + \sum_{e\in p_v} \frac{c_e}{N_e + 1} & > & \sum_{e\in p_u} \frac{c_e}{N_e + 1} \\
		\text{i.e., } \quad c_{uv} + \sum_{e\in p_v\setminus p_u} \frac{c_e}{N_e+1} + \sum_{e\in p_v\cap p_u} \frac{c_e}{N_e+1} & > & \sum_{e\in p_u\setminus p_v} \frac{c_e}{N_e + 1} + \sum_{e\in p_u\cap p_v} \frac{c_e}{N_e + 1}\\
		\text{i.e., } \quad c_{uv} + \sum_{e\in p_v\setminus p_u} \frac{c_e}{N_e + 1} & > & \sum_{e\in p_u\setminus p_v} \frac{c_e}{N_e + 1} \\
		\text{i.e., } \quad c_{uv} + \sum_{e\in p_v\setminus p_u} \frac{c_e}{N'_e + 1} & > & \sum_{e\in p_u\setminus p_v} \frac{c_e}{N'_e + 1} \\
		\text{i.e., } \quad c_{uv} + \sum_{e\in p_v\setminus p_u} \frac{c_e}{N'_e + 1} + \sum_{e\in p_v\cap p_u} \frac{c_e}{N'_e + 1} & > & \sum_{e\in p_u\setminus p_v} \frac{c_e}{N'_e + 1} + \sum_{e\in p_u\cap p_v} \frac{c_e}{N'_e + 1} \\
		\text{i.e., } \quad c_{uv} + \sum_{e\in p_v} \frac{c_e}{N'_e + 1} & > & \sum_{e\in p_u} \frac{c_e}{N'_e + 1}.
	\end{eqnarray*}
	Hence, $u$ moving to $v$ is not an improving move after $u$'s move to $x$.
\end{proof}

\begingroup
\def\thetheorem{\ref{lemma:two-non-leaf}}
\begin{claim}
  Let $T$ be a routing tree in a \nonleaf but not a \leaf state,
  arrived at by implementing one of the steps
  \eqref{move:amort}--\eqref{move:nonleaf} in algorithm \stm.  Let $u$ and $v$ be the
  non-leaf vertices charging the special cut $S^*$. Then either $u$ to
  $v$ or $v$ to $u$ is an improving move.
\end{claim}
\addtocounter{theorem}{-1}
\endgroup
\begin{proof}
 By the definition of the \nonleaf state, it must be the case that
 either $u$ or $v$ was the last to start a tree-follow move. Without
 loss of generality, say that $u$ was the last vertex to move. After
 $u$'s move, let $p_u$ and $p_v$ denote the routing paths of $u$ and
 $v$, where $(u, x)$ and $(v, y)$ are respectively the parent edges
 of $u$ and $v$. Then, we have $c_{ux}, c_{vy} \in
 [2^{i+2},2^{i+3})$. Moreover, $c_{uv} < 2^{i}$ since $u$ and $v$
 both belong to the cut $S^*$ at level $i$. Then, we have
\begin{equation}
\label{eq:cont}
	c_{ux}, c_{vy} > 4 c_{uv}.
\end{equation}	
We first claim that prior to $u$'s move to $x$, $u$ did not have an
improving move to $v$. Suppose, for contradiction, that $u$ did have
an improving move to $v$. Then, since $(u,v)$ is shorter than $(u,x)$,
and $u$ and $v$ are non-leaves, $u$'s potential move would have
triggered Steps~\eqref{move:amort}, \eqref{move:int2}, or
\eqref{move:nonleaf}. In each of these cases, $u$ would have preferred
the move to the closer vertex $v$ over the move to $x$.
It follows that $u$ to $v$ was not an improving move before $u$'s
move to $x$. Claim~\ref{lma:tree-follow} ensures that $u$ moving to $v$ is not
an improving move after $u$'s move to $x$ either. We therefore have,
\begin{align}
	c_{ux} + \sum_{e\in p_u \setminus (u, x)} \frac{c_e}{N_e}
	& \leq c_{uv} + \sum_{e\in p_v\setminus p_u} \frac{c_e}{N_e+1}
       + \sum_{e\in p_v\cap p_u} \frac{c_e}{N_e} \notag \\
\intertext{Here $N_e$ denotes the number of terminals using edge $e$ after $u$'s tree-move.}
	\text{i.e., } \quad c_{ux} + \sum_{e\in (p_u \setminus (u, x))\setminus p_v} \frac{c_e}{N_e} + \sum_{e\in p_{v}\cap p_u} \frac{c_e}{N_e}
	& \leq c_{uv} + \sum_{e\in p_v \setminus p_u}
       \frac{c_e}{N_e+1} + \sum_{e\in p_v\cap p_u} \frac{c_e}{N_e}
       \notag \\
	\text{i.e., } \quad c_{ux} + \sum_{e\in (p_u \setminus (u, x))\setminus p_v} \frac{c_e}{N_e}
	& \leq c_{uv} + \sum_{e\in p_v \setminus p_u}
       \frac{c_e}{N_e+1}. \label{eq:nonleaf1}
\end{align}
The rest of the proof is devoted to showing that $v$ moving to $u$ is
an improving move. Let us assume for the sake of contradiction that
$v$ moving to $u$ is not an improving move either. Then, we have
\begin{eqnarray}
	\sum_{e\in p_v} \frac{c_e}{N_e}
	& \leq & c_{uv} + \sum_{e\in p_u\setminus p_v} \frac{c_e}{N_e+1} + \sum_{e\in p_u\cap_v} \frac{c_e}{N_e} \nonumber\\
	\text{i.e., } \quad \sum_{e\in p_v\setminus p_u} \frac{c_e}{N_e} + \sum_{e\in p_v\cap p_u} \frac{c_e}{N_e}
	& \leq & c_{uv} + \sum_{e\in p_u\setminus p_v} \frac{c_e}{N_e+1} + \sum_{e\in p_u\cap_v} \frac{c_e}{N_e} \nonumber\\	
	\text{i.e., } \quad \sum_{e\in p_v\setminus p_u} \frac{c_e}{N_e}
	& \leq & c_{uv} + \sum_{e\in p_u\setminus p_v} \frac{c_e}{N_e+1}\label{eq:nonleaf2}
\end{eqnarray}
Adding the inequalities~\eqref{eq:nonleaf1} and \eqref{eq:nonleaf2}, and observing that $N_{ux}\geq 1$, we get
\begin{eqnarray*}
	c_{ux} + \sum_{e\in (p_u\setminus (u, x))\setminus p_v} \frac{c_e}{N_e} + \sum_{e\in p_v\setminus p_u} \frac{c_e}{N_e}
	& \leq & c_{uv} + \sum_{e\in p_v\setminus p_u} \frac{c_e}{N_e+1} + c_{uv} + \sum_{e\in p_u\setminus p_v} \frac{c_e}{N_e+1} \\
	\text{i.e., } \quad c_{ux} + \sum_{e\in (p_u\setminus (u, x))\setminus p_v} \frac{c_{e}}{N_e} + \sum_{e\in p_v\setminus p_u} \frac{c_e}{N_e}
	& \leq & 2 c_{uv} + \sum_{e\in p_v \setminus p_u} \frac{c_e}{N_e+1} + \frac{c_{ux}}{N_{ux} + 1} + \sum_{e\in (p_u \setminus p_v)\setminus (u, x)} \frac{c_e}{N_e+1} \\	
	\text{i.e., } \quad c_{ux} &\leq & 2 c_{uv} + \frac{c_{ux}}{2} \\
	\text{i.e., } \quad c_{ux} & \leq & 4 c_{uv},
\end{eqnarray*}
which contradicts inequality~(\ref{eq:cont}).
Thus, it must be the case that $v$ moving to $u$ is an improving move.
\end{proof}

\begingroup
\def\thetheorem{\ref{lma:leaf}}
\begin{lemma}
	Suppose a set of new terminals $U_t$ arrive in epoch $t$ when the routing paths of the
	existing terminals are in an equilibrium state. Then, for each new terminal $u\in U_t$,
	the chosen routing path $p_u$ comprises a single edge $(u, v)$
	connecting $u$ to an existing vertex $v$ in the current routing tree $T$, and then
	following the unique path in $T$ from $v$ to $r$.
\end{lemma}
\addtocounter{theorem}{-1}
\endgroup
\begin{proof}
	The lemma has two parts:
	\begin{itemize}
		\item $p_u$ has a single edge $(u, v)$ connecting $u$ to $v$, and
		\item the path $p_u$ does not deviate from $T$ between $v$ and $r$.
	\end{itemize}
	The first statement is a direct consequence of our tie-breaking
       rule for best response paths: $u$ is the only terminal using
       the portion of $p_u$ from $u$ to $T$; if this segment consists
       of a multi-hop path from $u$ to some vertex $v\in T$,
       short-cutting this segment and using the direct $(u,v)$ edge
       instead is potentially cheaper and has fewer edges with only
       $u$ using them.

	For the sake of contradiction, suppose the second statement is
       false. Then we claim that the vertex $v$ has an improving
       move. This contradicts the fact that the routing paths are in
       equilibrium when terminal $u$ arrives. To prove the claim,
       suppose first that $v$ is a terminal. Let $p_v$ denote the
       path along $T$ from $v$ to the root, and let $q_v$ denote the
       segment of $p_u$ from $v$ to the root. Let $E_v$ be the set of
       edges in $p_v\setminus q_v$, and $E_u$ the set of edges in
       $q_v\setminus p_v$. Let $N_e$ denote the number of terminals
       using edge $e$ prior to any arrivals in this epoch. Since $u$
       follows its best response path, we have that
       \begin{align*}
         \sum_{e\in E_u} \frac{c_e}{N_e+1} \le \sum_{e\in E_v} \frac{c_e}{N_e+1}.
       \end{align*}
       The cost share of $v$ over edges in $E_v$ is $\sum_{e\in E_v}
       \frac{c_e}{N_e}$, whereas over edges in $E_u$ if $v$ were to
       switch to taking the path $q_v$ would be $\sum_{e\in E_u}
       \frac{c_e}{N_e+1}$. From the above inequality, it follows that
       \begin{align*}
         \sum_{e\in E_u} \frac{c_e}{N_e+1} < \sum_{e\in E_v} \frac{c_e}{N_e},
       \end{align*}
       which implies that $v$'s cost share would improve strictly by
       switching from $p_v$ to $q_v$. When $v$ is not a terminal, but
       is on the current path of some other terminal $w$, an
       identical argument shows that $w$ (and therefore $v$) has an
       improving move.
\end{proof}

\begingroup
\def\thetheorem{\ref{lma:dep+arr}}
\begin{lemma}
 After the arrival or departure of a set of terminals in an
  \equilibrium state, the routing tree $T$ remains in a \leaf state.
\end{lemma}
\addtocounter{theorem}{-1}
\endgroup
\begin{proof}
  For an arrival event, this is a direct consequence of
  Lemma~\ref{lma:leaf}. Since every new terminal is a leaf, there is
  at most one non-leaf vertex charging every cut. After a set of
  terminal departures, the routing tree clearly remains in a
  \amortized state since charges to dual cuts can only decrease.
\end{proof}

\begingroup
\def\thetheorem{\ref{lma:single-edge}}
\begin{lemma}
	If the routing tree is not in equilibrium, then at least one improving
	tree-follow move exists.
\end{lemma}
\addtocounter{theorem}{-1}
\endgroup
\begin{proof}
  Let the current routing path of every terminal $x$ be denoted $p_x$
  and the union of these paths be tree $T$. For notational
  convenience, we also denote the path in the routing tree from a
  non-terminal vertex $x$ to $r$ by $p_x$. We first identify a vertex
  $w$ which has an improving path $q_w$ that contains exactly one arc
  not in $T$. Since $T$ is not an equilibrium, there exists a terminal
  $x$ whose best response path $p_{x}^*$ differs from $p_x$. Let
  $(u,v)$ be the edge that is closest to $r$ on $p^*_x$ and is not in
  $T$, with $v$ closer to $r$ than $u$. Let $w$ be any terminal in the
  subtree of $T$ rooted at $u$. We claim the path $q_w$ formed by
  taking the subpath in $T$ from $w$ to $u$ and then the subpath of
  $p^*_x$ from $u$ to $r$ is an improving path for $w$. Indeed, if its
  shared cost is at least the shared cost of $p_w$, we can find a path
  for $x$ whose shared cost is at most the shared cost of
  $p^*_x$. Consider the path $q_x$ for $x$ formed by taking the
  subpath $p^*_x$ from $x$ to $u$ and then taking the subpath from $u$
  to $r$. The difference in shared cost for $x$ between $q_x$ and
  $p^*_x$ equals the difference in shared cost for $w$ in $p_w$ and
  $q_w$. Due to our tie-breaking rule, the best response path $p^*_x$
  must be strictly better than $q_x$ since $q_x$ has fewer new
  edges. Thus $q_w$ must be strictly better than $p_w$ as desired.

  Now let $T'=T\setminus (u,\text{parent}(u))\cup (u,v)$ where $\text{parent}(u)$ is the parent of $u$ in $T$. We now give a series of improving moves which give us the routing tree $T'$. Order the terminals in subtree of $T$ rooted at $u$. For each such terminal $w$ change its path from $p_w$ to $q_w$ as defined above. It is clear that the union of all routing paths is exactly $T'$. It remains to show that they are all improving paths for their respective terminals. From the above argument, for the first terminal $w_1$, $q_{w_1}$ is clearly an improving path over $p_{w_1}$ in $T$. For any latter terminal $w$ in the sequence, the shared cost of $q_w$ has only reduced as compared to its shared cost in the solution $T$ since more terminals are using the edges on the subpath from $u$ to $x$. Moreover, the shared cost of $p_w$ has only increased as compared to its shared in $T$ as fewer terminals are using the subpath of $w$ from $u$ to $r$. The subpath from $w$ to $u$ remains identical and same number of terminals keep using it and therefore, $q_w$ remains an improving path in the intermediate solution as well for each of the terminals in the sequence.
\end{proof}

\section{Constructing the Dual in an Online Fashion}
\label{sec:online}

The classification of tree routings described in
Section~\ref{sec:classification} as well as the description of
algorithm \stm relies on the knowledge of the dual family $\Pi$
defined in Section~\ref{sec:classification} to which we charge the
cost of our solution. If the underlying graph $G$ is known in advance,
there are standard techniques for constructing a family of duals with
the desired properties. In our setting, the underlying graph may not
be known in advance, and may instead be revealed over time as
terminals arrive. We now describe a simple greedy procedure for
constructing the dual family in an online fashion.

Recall that the graph $G$ is complete and edge lengths form a metric,
i.e. they satisfy the triangle inequality. We use $c(u,v)$ to denote
the length of the edge between vertices $u$ and $v$. The graph is
revealed vertex by vertex, and every time a vertex is added, all edges
between that vertex and previously added vertices are revealed. The
level $j$ dual for any integer $j$ is constructed as follows. At any
point of time, we have some number of components $S_1, \cdots, S_m$ in
the dual, with centers $s_1, \cdots, s_m$, respectively. At the
beginning when the graph contains a single vertex, we have a single
component with that vertex as its center. When the next vertex, say
$v$, arrives, if there exists a center $s_i$ with $c(s_i,v)< 2^{j-1}$,
we add $v$ to the $i$th component $S_i$. Otherwise, we create a new
component $S_{m+1}$ with center $s_{m+1} = v$. By construction, it
holds that every vertex in component $S_i$ is at a distance less than
$2^{j-1}$ from its center $s_i$, and therefore, the diameter of the
component is less than $2^j$. Also, by construction, the distance
between any two centers is at least $2^{j-1}$. Therefore, the
constructed dual satisfies the properties listed in
Section~\ref{sec:classification}.

\section{Bad Examples}\label{app:examples}

\begin{figure}
\begin{minipage}[t][6cm]{0.5\linewidth}
\subfloat[Example 1]{\includegraphics[width=3cm,height=7cm]{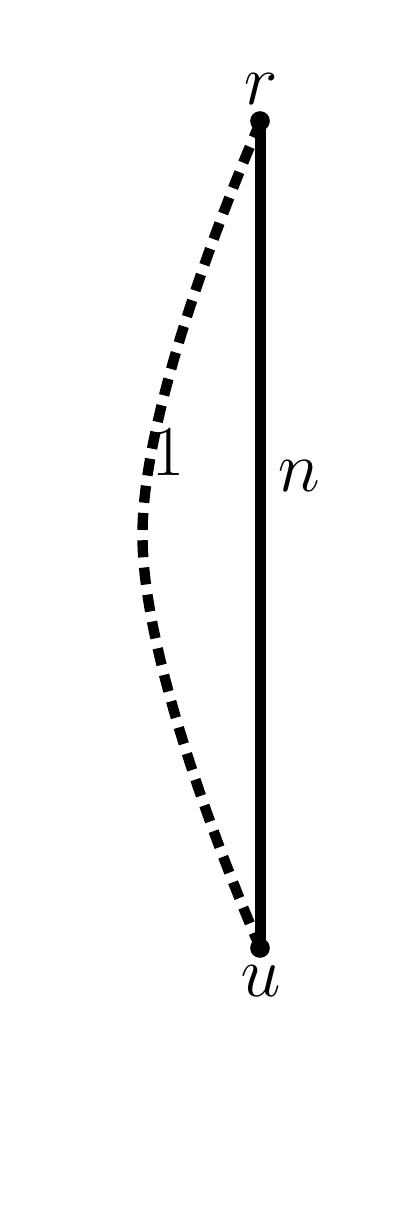}}
\end{minipage}%
\begin{minipage}[t][6cm]{0.5\linewidth}
\subfloat[Example 2]{\includegraphics[width=3cm,height=7cm]{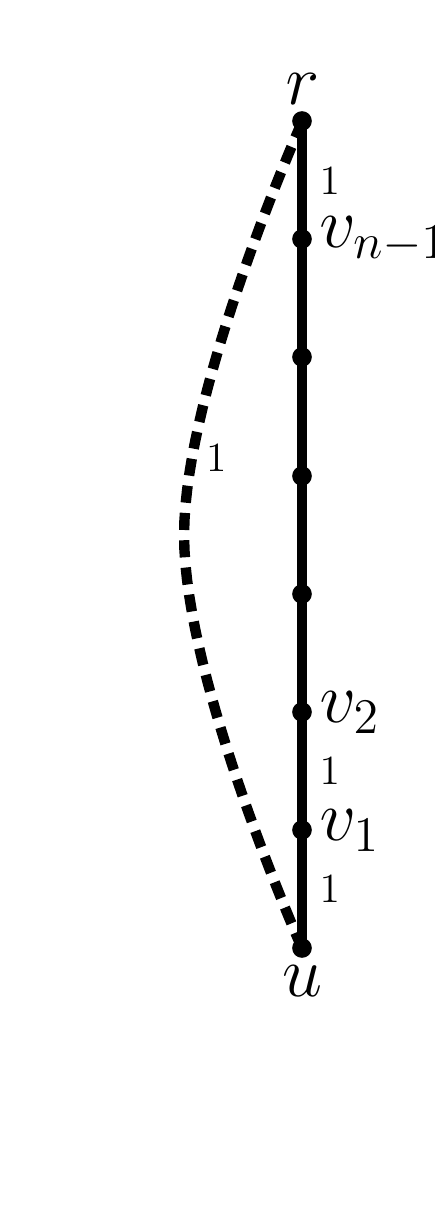}}
\end{minipage}
\\\\\\\\
\caption{Bad Examples for Price of Anarchy and comparison with Minimum Steiner Tree}\label{fig:poa}
\end{figure}

In Figure~\ref{fig:poa}(a) we give an instance from \cite{AnshelevichDKTWR08} where the price of anarchy is arbitrary large. In the above example, there are $n+1$ agents at $u$. If the solution picks the edge of weight $n$ instead of the edge of weight $1$, a simple check shows that it is still in equilibrium. Since the optimal solution picks the edge of weight $1$ instead of weight $n$, the price of anarchy is $n$. 

In Figure~\ref{fig:poa}(b), we give an instance where the natural dynamics leads to solution that is much more expensive than the minimum Steiner tree. Consider the following sequence of arrivals where each agent picks the best response path on arrival. First $n$ agents arrive on $v_{n-1}$, then $n$ agents arrive on $v_{n-2}$, and so on. Finally $n$ agents arrive on $u$. Observe that in each phase of $n$ arrivals, the best response dynamics introduces the edge $(v_{i},v_{i+1})$ and thus the solution at the end is the long path from $u$ to $r$. Now, all the agents, except the $n$ agents at $u$, depart. Observe that the solution is still in equilibrium since it is identical to the price of anarchy solution in Figure~\ref{fig:poa}. But the weight of minimum Steiner tree is $1$ for the agents which survive. This shows that the dynamics can lead to a much costlier solution as compared to the minimum Steiner tree. A more apt comparison is to the cost of the minimum spanning tree over all of the arriving clients. In this case, the minimum spanning tree costs $n$ which is exactly our solution.

\pagebreak
\section{Illustration of a Tree-Follow Move}
\label{sec:tree-move-figs}
\begin{figure}[!h]
\begin{minipage}[t][6cm]{0.5\linewidth}
\subfloat[Solution A]{\includegraphics[width=8cm,height=10cm]{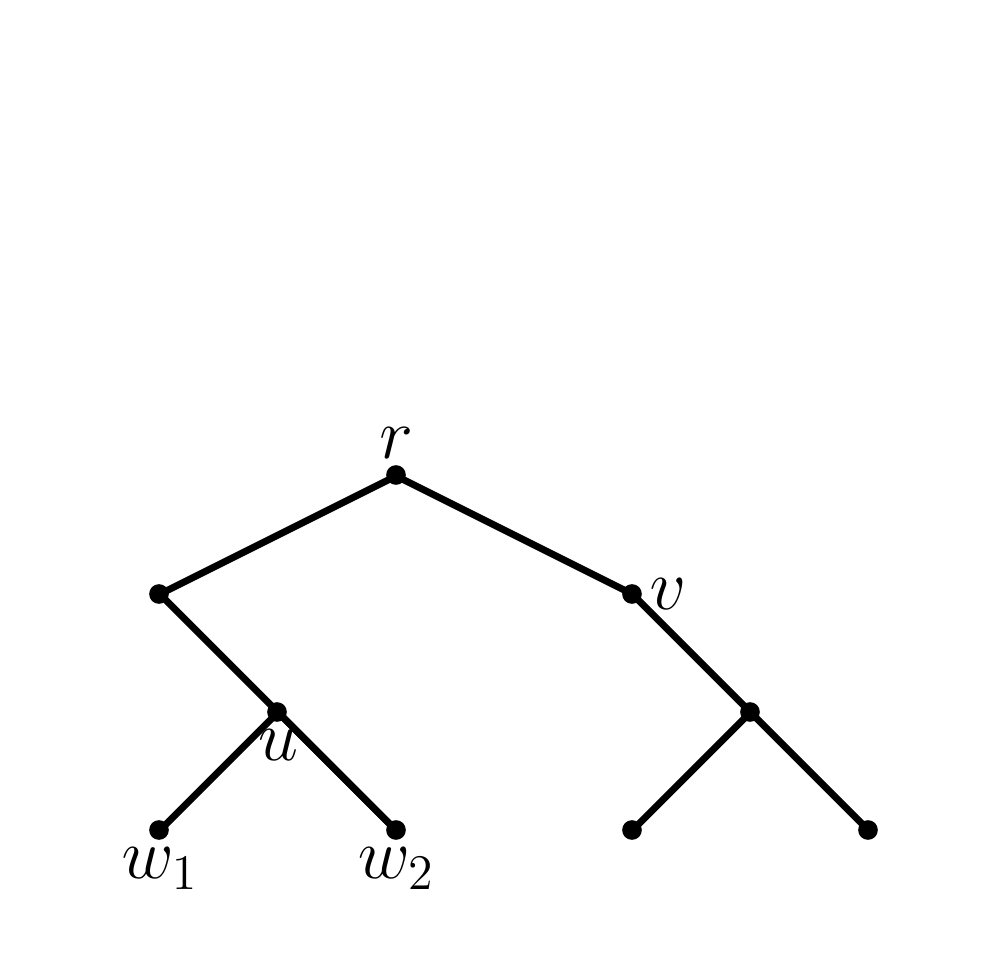}}
\end{minipage}%
\begin{minipage}[t][6cm]{0.5\linewidth}
\subfloat[Solution B]{\includegraphics[width=8cm,height=10cm]{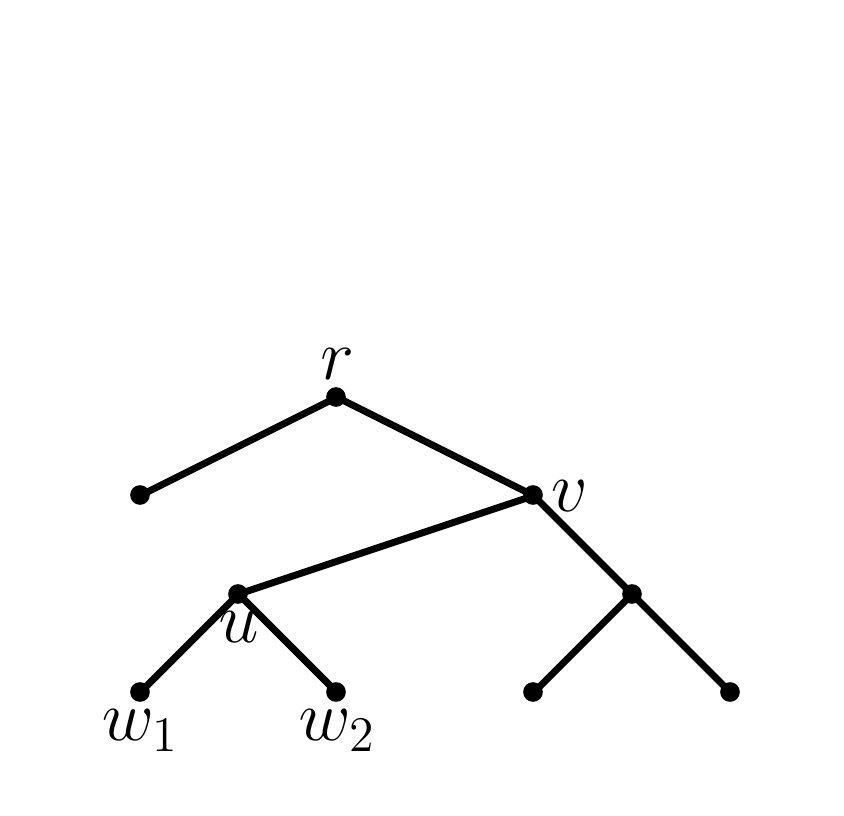}}
\end{minipage}
\\\\\\\\\\\\\\\\\\\\
\caption{Tree Move from $u$ to $v$ changes the solution from A to B.}\label{foo1}
\end{figure}

\begin{figure}[!h]
\begin{minipage}[t][4cm]{0.25\linewidth}
\subfloat[Solution A]{\includegraphics[width=4cm,height=6cm]{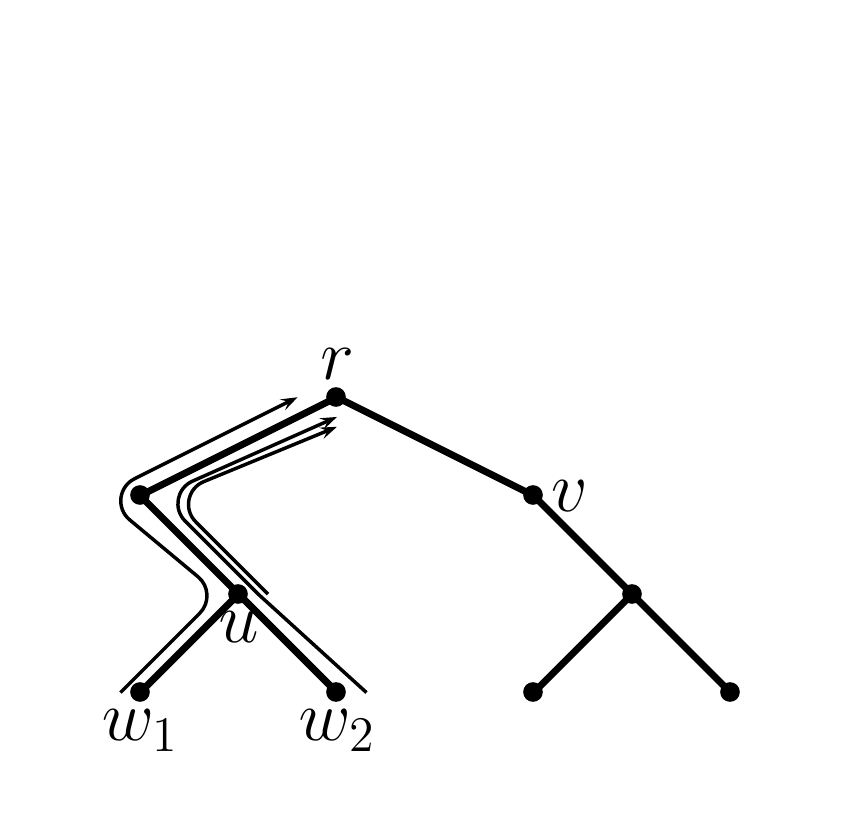}}
\end{minipage}%
\begin{minipage}[t][4cm]{0.25\linewidth}
\subfloat[Solution Int1]{\includegraphics[width=4cm,height=6cm]{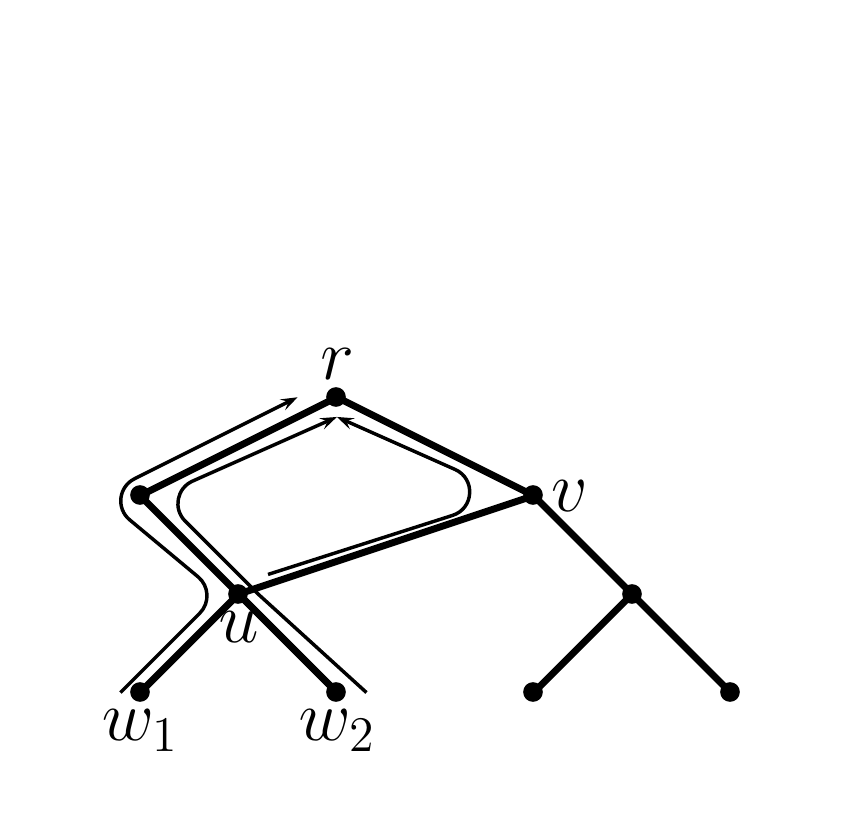}}
\end{minipage}
\begin{minipage}[t][4cm]{0.20\linewidth}
\subfloat[Solution Int2]{\includegraphics[width=4cm,height=6cm]{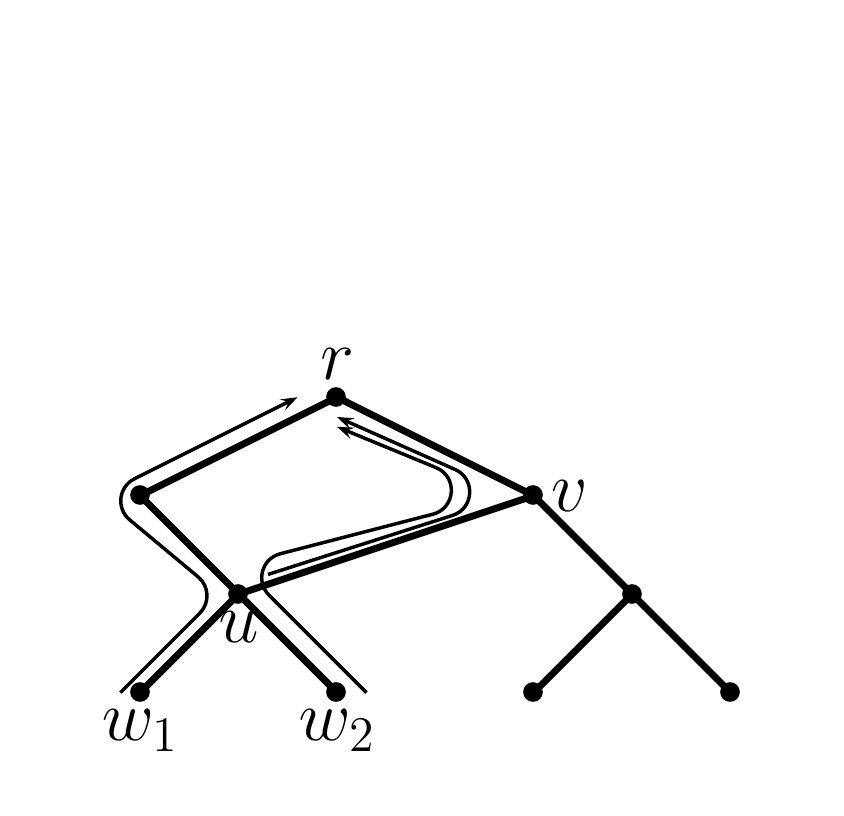}}
\end{minipage}
\begin{minipage}[t][4cm]{0.20\linewidth}
\subfloat[Solution B]{\includegraphics[width=4cm,height=6cm]{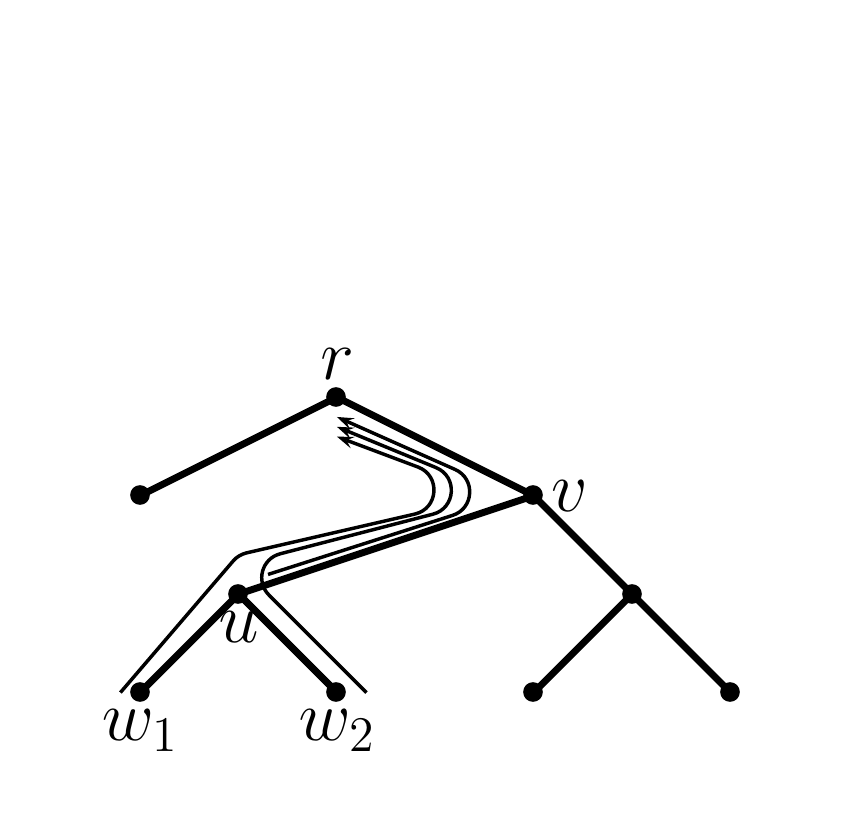}}
\end{minipage}
\\\\\\\\\\\\
\caption{Implementing the tree-follow move by individual moves. In Solution A, the current paths for $u$ and its children $w_1$ and $w_2$ are shown. In Solution Int1, $u$ changes its path and introduces a new edge $(u,v)$. Vertices $w_1$ and $w_2$ follow $u$'s path in Solution Int2 and D, respectively.}\label{foo2}
\end{figure}

\end{document}